\documentclass{optica-article}

\journal{opticajournal} % for journals or Optica Open

\articletype{Research Article}

\usepackage{lineno}
\usepackage{xr}
\usepackage{soul}
\usepackage{amsmath}
\usepackage{graphicx}
\usepackage{epstopdf}
\usepackage{subfig}
\usepackage{multirow}
\usepackage{url}
\usepackage{float}
\usepackage{bm}
\usepackage{ifthen}
\usepackage{braket}

\usepackage{siunitx} %able to align table columns by decimal point

\usepackage[absolute]{textpos}
\usepackage{xcolor}
\usepackage{placeins}

\begin{document}

\title{Two-mode squeezing over deployed fiber coexisting with conventional communications}

\author{Joseph C. Chapman,\authormark{1,*} Alexander Miloshevsky,\authormark{1} Hsuan-Hao Lu,\authormark{1} Nageswara Rao,\authormark{2} Muneer Alshowkan,\authormark{1} and Nicholas A. Peters\authormark{1}}

\address{\authormark{1} Quantum Information Science Section, Oak Ridge National Laboratory, Oak Ridge, TN 37831, USA\\
\authormark{2} Advanced Computing Methods for Engineered Systems Section, Oak Ridge National Laboratory, Oak Ridge, TN 37831, USA}

\email{\authormark{*}chapmanjc@ornl.gov} %% email address is required

%\author{Joseph C. Chapman}
%\email{chapmanjc@ornl.gov}
%\affiliation{Quantum Information Science Section, Oak Ridge National Laboratory, Oak Ridge, TN 37831}
%\author{Alexander Miloshevsky}
%\affiliation{Quantum Information Science Section, Oak Ridge National Laboratory, Oak Ridge, TN 37831}
%\author{Hsuan-Hao Lu}
%\affiliation{Quantum Information Science Section, Oak Ridge National Laboratory, Oak Ridge, TN 37831}
%\author{Nageswara Rao}
%\affiliation{Advanced Computing Methods for Engineered Systems Section, Oak Ridge National Laboratory, Oak Ridge, TN 37831}
%\author{Muneer Alshowkan}
%\affiliation{Quantum Information Science Section, Oak Ridge National Laboratory, Oak Ridge, TN 37831}
%\author{Nicholas A. Peters}
%\affiliation{Quantum Information Science Section, Oak Ridge National Laboratory, Oak Ridge, TN 37831}

\begin{abstract*}
Squeezed light is a crucial resource for continuous-variable (CV) quantum information science. Distributed multi-mode squeezing is critical for enabling CV quantum networks and distributed quantum sensing. To date, multi-mode squeezing measured by homodyne detection has been limited to single-room experiments without coexisting classical signals, i.e., on ``dark'' fiber. Here, after distribution through separate fiber spools (5~km), $-0.9\pm0.1$-dB coexistent two-mode squeezing is measured. Moreover, after distribution through separate deployed campus fibers (about 250~m and 1.2~km), $-0.5\pm0.1$-dB coexistent two-mode squeezing is measured. Prior to distribution, the squeezed modes are each frequency multiplexed with several classical signals---including the local oscillator and conventional network signals---demonstrating that the squeezed modes do not need dedicated dark fiber. After distribution, joint two-mode squeezing is measured and recorded for post-processing using triggered homodyne detection in separate locations. This demonstration enables future applications in quantum networks and quantum sensing that rely on distributed multi-mode squeezing.
\end{abstract*}
%\maketitle

\begin{textblock}{12.13}(1.94,15)
\noindent\fontsize{7}{7}\selectfont \textcolor{black!30}{This manuscript has been co-authored by UT-Battelle, LLC, under contract DE-AC05-00OR22725 with the US Department of Energy (DOE). The US government retains and the publisher, by accepting the article for publication, acknowledges that the US government retains a nonexclusive, paid-up, irrevocable, worldwide license to publish or reproduce the published form of this manuscript, or allow others to do so, for US government purposes. DOE will provide public access to these results of federally sponsored research in accordance with the DOE Public Access Plan (http://energy.gov/downloads/doe-public-access-plan).}
\end{textblock}

%%%%%%%%%%%%%%%%%%%%%%%%%%  body  %%%%%%%%%%%%%%%%%%%%%%%%%%
\section{Introduction} 
%Expand introduction
Two-mode squeezed vacuum is a necessary resource for continuous-variable (CV) quantum repeaters~\cite{PhysRevA.98.032335,Wu_2022,PhysRevA.106.052604}. Multi-mode squeezing enables distributed CV entanglement and distributed quantum sensing~\cite{Guo2020,PhysRevResearch.2.023030,Zhang_2021}; where, the latter still has a potential sensitivity advantage even in the presence of significant loss~\cite{Guo2020}. In these proposals and prior experiments, multi-mode squeezing is primarily created by single-mode squeezers and beamsplitter networks. While free-space squeezing experiments have been the norm, fully fiber-based demonstrations have been successful~\cite{Kaiser:16}.  Furthermore, higher performance has been achieved by combining micro-free-space optical systems~\cite{doi:10.1063/5.0063118}. These demonstrations have been enabled by the efficient single-pass squeezing generated in quasi-phase-matched lithium niobate waveguides~\cite{Serkland:95,Kanter:02}. Moreover, state-of-the-art two-mode squeezed-light fiber transmission has recently been demonstrated in a single-room utilizing dark fiber~\cite{doi:10.1126/sciadv.aas9401,Liu:18}. However, dark fiber can have high cost and limited availability; thus, coexistence of quantum and classical communications is desirable. In addition, demonstrations to date using multi-mode squeezing and homodyne detection have been limited to co-located joint homodyne detection. Co-located detection is not practical when distributing each squeezed mode to a different distant network node---which practical protocol implementations require.

%Homodyne detection of multi-mode squeezing was done with a local oscillator (LO) typically multiplexed using polarization and/or time degrees of freedom; such multiplexing is practical with low excess-loss in free space and when beginning with single-mode squeezing.  However, to achieve low-loss high-isolation multiplexing in optical fiber and take better advantage of the fiber's bandwidth with non-degenerate multi-mode squeezing, frequency multiplexing, using  ubiquitous wavelength-division multiplexing technology, is advantageous. 
% While this multiplexing scheme is practical and low-loss in free space, it is not readily compatible with the conventional networking frequency multiplexing paradigm in fiber--which offers lower-loss solutions due to the vast investments in wavelength-division multiplexing technology.

%Expand ``in this work statement''
To advance the state-of-the-art for practical distributed use, we begin by demonstrating fiber-based two-mode squeezing,  in the telecommunications C-band (about 1530-1565~nm).   The squeezed light is frequency multiplexed with multiple classical signals---including the local oscillator (LO) and conventional optical-communications signals---with spacing as low as 0.8~nm (100~GHz) using conventional dense wavelength-division multiplexing. Although (de)multiplexing adds some insertion loss, the squeezing is not measurably degraded by the classical light sharing the fiber, which stands in stark contrast to the degradation due to Raman scattering for discrete variables~\cite{Peters_2009}. Moreover, frequency multiplexing of the LO and squeezed light avoids excess noise from guided acoustic wave Brillouin scattering which has degraded two-mode squeezing in a previous demonstration using polarization multiplexing~\cite{doi:10.1126/sciadv.aas9401}. Coexistent two-mode squeezing is demonstrated using 10-km of fiber on spools in the same lab and, more importantly, across a 3-node campus network using deployed fiber. After fiber transmission and demultiplexing, the LO undergoes synchronized wavelength conversion to the pair of squeezed-mode wavelengths.  Next, distributed detection records the two-mode squeezing for post-processing via triggered homodyne detection. Our demonstration of two-mode squeezing distributed on deployed fiber while coexisting with conventional networking signals combined with distributed joint detection lays a foundation for future CV quantum networking applications.

\begin{figure}%[t!]%scale=0.35width=5.5in,height=1.65in
\centerline{\includegraphics[width=1\textwidth]{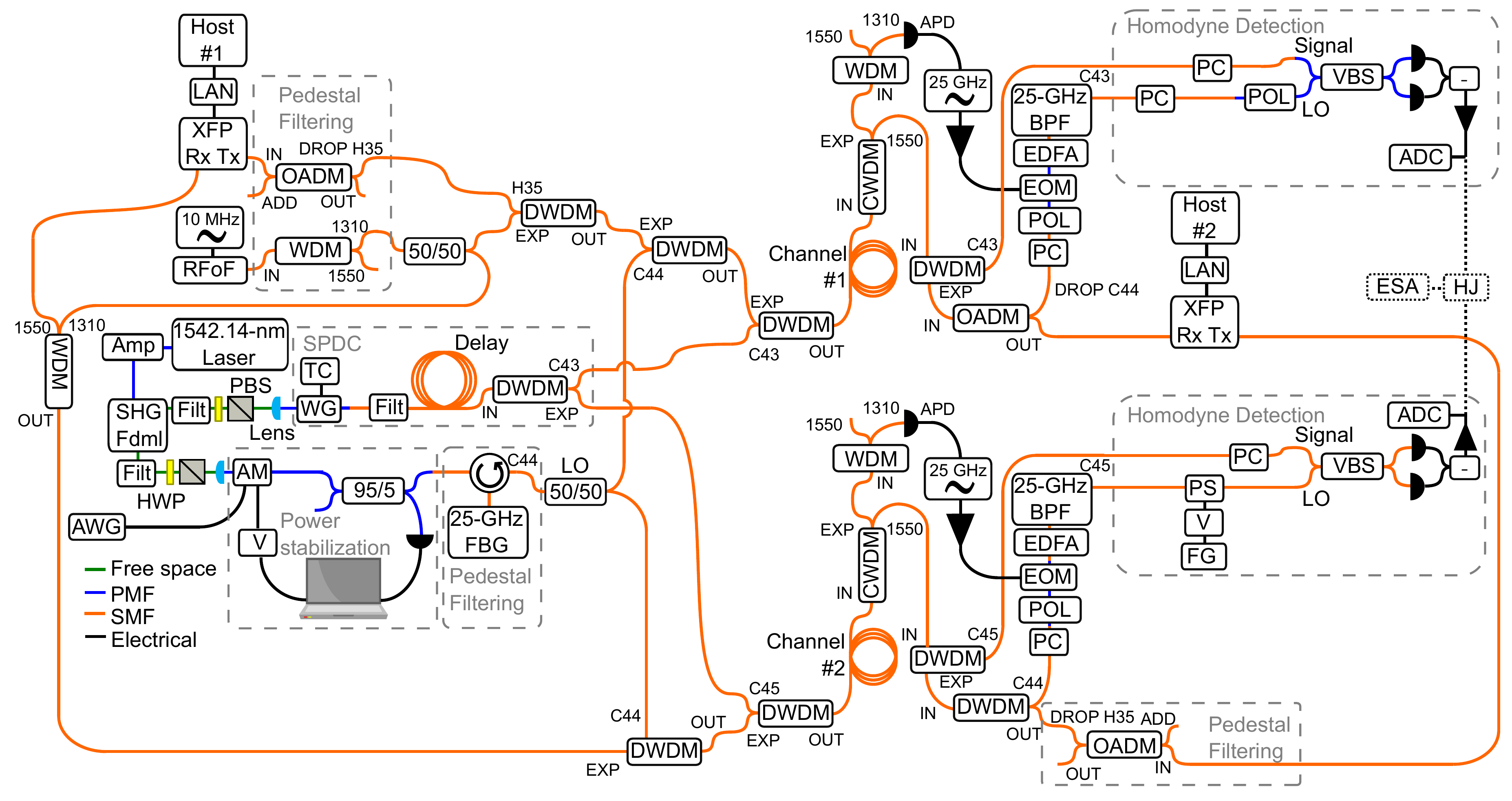}}
\caption{Two-mode squeezing generation and measurement setup. The 1542.14-nm laser is amplified and used to create the 771-nm pump. The pump produces squeezed vacuum (centered at 1542.14 nm) in the waveguide [Fig.~\ref{fig:SPDC}]. The two modes of interest (C43 and C45) are split using a DWDM and combined with multiple classical signals [Fig.~\ref{fig:CHspec}]. The leftover 1542.14-nm laser is used as the phase reference and is multiplexed with a conventional networking signal and 10-MHz sideband reference RFoF signal before being multiplexed with the squeezed light. After transmission through the channel, the signals are demultiplexed on each side. The phase and sideband references are used to create the local oscillator at each detection node for homodyne detection [Fig.~\ref{fig:sidebandsOSA}]. A trigger pulse is applied using an AWG and AM to the phase reference which propagates through the system to each detection node. The ADCs are triggered by this signal to acquire quadrature data which is post processed later. Alternatively, when the whole experiment is in a single location, the homodyne detector outputs can be electrically combined directly using a hybrid junction and analyzed in real-time using an ESA. ADC = analog-to-digital converter, AM = amplitude modulator, AWG = arbitrary waveform generator, BPF = band-pass filter, EDFA = erbium-doped fiber amplifier, EOM = electro-optic modulator, EXP = express channel, FBG = fiber-bragg grating, Fdml = fundamental, FG = function generator, Filt = filter, HWP = half-wave plate, LAN = local-area network, OADM = optical add-drop multiplexer, PC = polarization controller, POL = polarizer, PBS = polarizing beamsplitter, PS = phase shifter, TC = temperature controller, V = voltage supply, VBS = variable beamsplitter, WDM = wavelength-division multiplexing, WG = waveguide, XFP = 10-Gbps small-form-factor pluggable transceiver.}
\label{fig:setup}
\end{figure}

\section{Methods}

\begin{figure}%[t!]
\centerline{\includegraphics[width=0.5\textwidth]{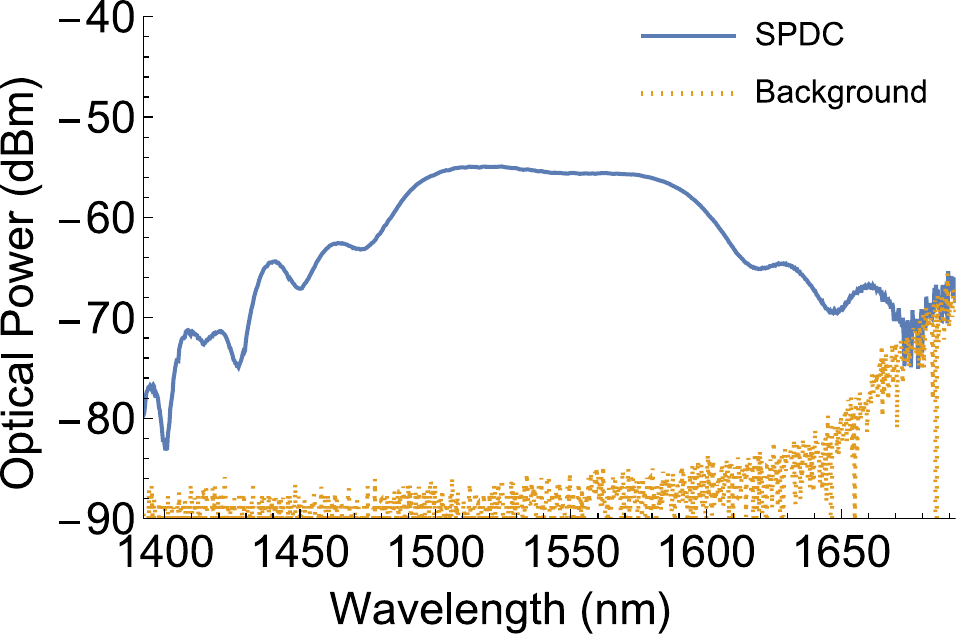}}
\caption{Waveguide spectra. Spectra of spontaneous parametric downconversion from PPLN waveguide. Spectra measured using a resolution bandwidth of 0.5~nm.}
\label{fig:SPDC} %fig:SPDC -> fig:OEmethods(a)
\end{figure}

Broadband squeezing is produced by fiber-coupled continuous-wave 771-nm second-harmonic generation (SHG) pumping a fiber-coupled 2.5-cm type-0 periodically poled lithium-niobate (PPLN) ridge waveguide (AdvR, Inc.)~[Fig.~\ref{fig:setup}]. The waveguide couples the squeezed light into the output fiber with 60\% coupling efficiency; this can be improved with further engineering~\cite{doi:10.1063/5.0063118}.  The waveguide spontaneous-parametric-downconversion bandwidth is pretty flat over 60-nm [Fig.~\ref{fig:SPDC}]; broadband squeezing has been measured over wider bandwidths in similar devices~\cite{doi:10.1063/5.0063118}. Two-mode squeezing of 1542.94~nm (channel 43, C43) and 1541.35~nm (channel 45, C45) is the focus of this work. Each squeezed mode  is distributed into a separate fiber using a dense wavelength-division multiplexing (DWDM) filter (AC Photonics). Use of DWDM filters is enabled by the SHG pump being aligned onto channel 44 (C44) of the ITU grid (1542.14 nm). Further laser and waveguide details are provided in \ref{app:lwsubsys}.

%\begin{figure}%[t!]
%\centerline{\includegraphics[width=1\textwidth]{OEmethodsfig}}
%\caption{(a) Spectra of spontaneous parametric downconversion from PPLN waveguide. Spectra measured using a resolution bandwidth of 0.5~nm. (b) Individual spectra of the signals before pedestal filtering and multiplexing, except the squeezed light went through multiplexing due to fusion spliced fibers. (c) Individual spectra after pedestal filtering and multiplexing. Measured at the input to Channel 1 after multiplexing all signals on the C43 side. Spectra measured with an optical spectrum analyzer, using a resolution bandwidth of 0.02~nm and an in-line calibrated 0.1-nm notch filter. (d) Distributed two-mode squeezing over deployed campus fiber. The source is in Building A. Channel \#1 (\#2) going to building B (C) is about 1200~m (250~m) with link loss of 4.6$\pm0.1$~dB (2$\pm0.1$~dB). Map Credit~\cite{gmaps}. (e) Spectra of local oscillator wavelength conversion from C44 to C43 using an electro-optic phase modulator to create optical sidebands which are then amplified and filtered. Spectra measured using a resolution bandwidth of 0.02~nm. (f) Spectral measurement of the demultiplexed signal going into the C43 homodyne detector. We measured the spectra for 0-km and 5-km with the squeezed-light blocked and 0-km with the squeezed light. We also include the measured noise-equivalent power at C43 using the ESA. Spectra measured with an optical spectrum analyzer, using a resolution bandwidth of 0.02~nm.}
%\label{fig:OEmethods}
%\end{figure}

\begin{figure}%[t!]
\centerline{\includegraphics[width=1\textwidth]{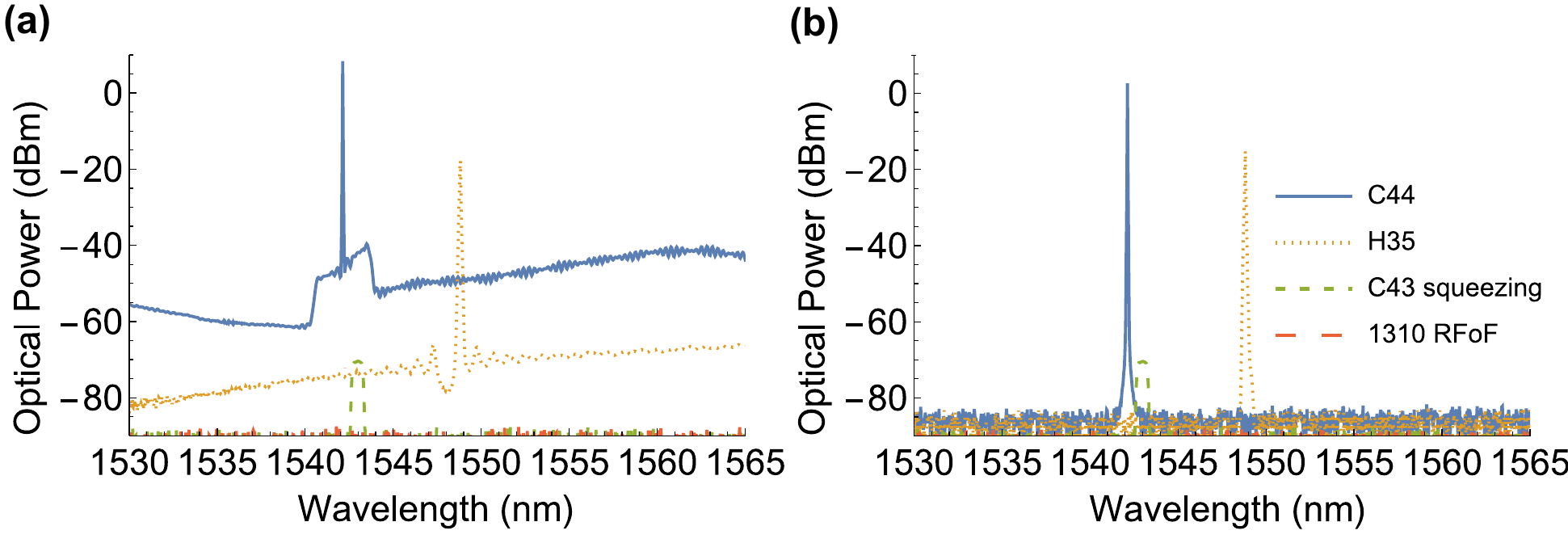}}
\caption{Multiplexed spectra in C-band. (a) Individual spectra of the signals before pedestal filtering and multiplexing, except the squeezed light went through multiplexing due to fusion spliced fibers. (b) Individual spectra after pedestal filtering and multiplexing. Measured at the input to Channel 1 after multiplexing all signals on the C43 side. Spectra measured with an optical spectrum analyzer, using a resolution bandwidth of 0.02~nm and an in-line calibrated 0.1-nm notch filter.}
\label{fig:CHspec} %fig:CHspec -> fig:OEmethods(b-c)
\end{figure}

The squeezed modes are each frequency multiplexed with several other signals before the fiber channels [see Fig.~\ref{fig:CHspec}]. The squeezed light is multiplexed with the LO at C44 and a 1310-nm radio-frequency over fiber (RFoF) signal using wavelength-division multiplexing filters, primarily DWDM filters (AC Photonics). The C43 squeezed mode is also multiplexed with a co-propagating conventional telecom signal (Finisar 10GigE XFP) at 1549~nm (H35); concurrently, the C45 mode is multiplexed with a counter-propagating classical signal.
The conventional signals are generated to exchange ethernet packets between the two 10GigE XFPs. The conventional traffic is generated by the iperf3 tool using TCP between the two hosts.  Further classical networking  details are in \ref{app:conventraff}.

Before multiplexing the classical signals with the squeezed light, those signals are filtered to remove any stray light at the squeezing wavelengths (C43 and C45), e.g., from the laser pedestal [see Fig.~\ref{fig:CHspec}]. The C44 LO is narrowly filtered, with low insertion loss, using a fiber-optic circulator (General Photonics)  %CIR-15-SM-90-FC/APC) 
and a 25-GHz fiber Bragg grating (O/E Land%OEDWDM-025)
). The H35 and 1310-nm signals are filtered using DWDM %(AC Photonics DWDM1H35102A221066)
and HWDM %(AC Photonics HWDM5531102A220511) 
filters (AC Photonics). For accurate pedestal spectral measurements [Fig.~\ref{fig:CHspec}], details are provided in \ref{app:lwsubsys}.

The (de)multiplexing components and all other components in the squeezed-light path should introduce the lowest loss possible for maximal squeezing. The components from the input of the pump-rejection (directly after the waveguide) to the input of the homodyne detector (with no added channel fiber) have a total insertion loss for each squeezed-light mode C43 (C45) of 1.5 (2.3)$\pm0.1$~dB. This insertion loss is achieved by fusion splicing most connections in the path of the squeezed light; those not spliced (for practical reasons) are the connections between the waveguide and pump-rejection filter, the connections at the beginning and end of the channel fiber, and the connection at the signal input to the homodyne detectors. 

\begin{figure}%[t!]
\centerline{\includegraphics[width=0.8\textwidth]{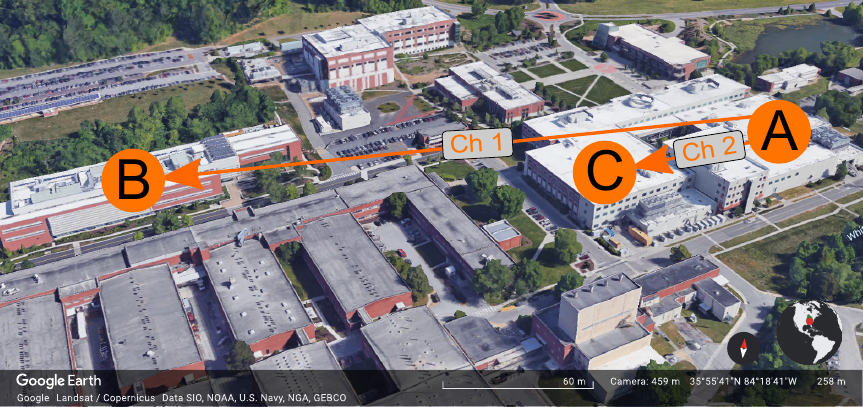}}
\caption{Distributed two-mode squeezing over deployed campus fiber. The source is in Building A. Channel \#1 (\#2) going to building B (C) is about 1200~m (250~m) with link loss of 4.6$\pm0.1$~dB (2$\pm0.1$~dB). Map Credit~\cite{gmaps}.}
\label{fig:QLAN} %fig:QLAN -> fig:OEmethods(d)
\end{figure}

Squeezing is measured using no added fiber and also using two 5-km spools (Corning SMF-28e) for channels. The 5-km spool used in Channel 1(2) has insertion loss of 1.1(1.2)$\pm0.1$dB. Furthermore, some measurements use deployed campus fiber [see Fig.~\ref{fig:QLAN}]. For these measurements, the source is in Building A, while the C43 (C45) demultiplexing and homodyne detection is in Building B (C); the fiber distance between A and B (C) is about 1200~m (250~m) with loss of 4.6$\pm0.1$~dB (2$\pm0.1$~dB).

\begin{figure}%[t!]
\centerline{\includegraphics[width=0.5\textwidth]{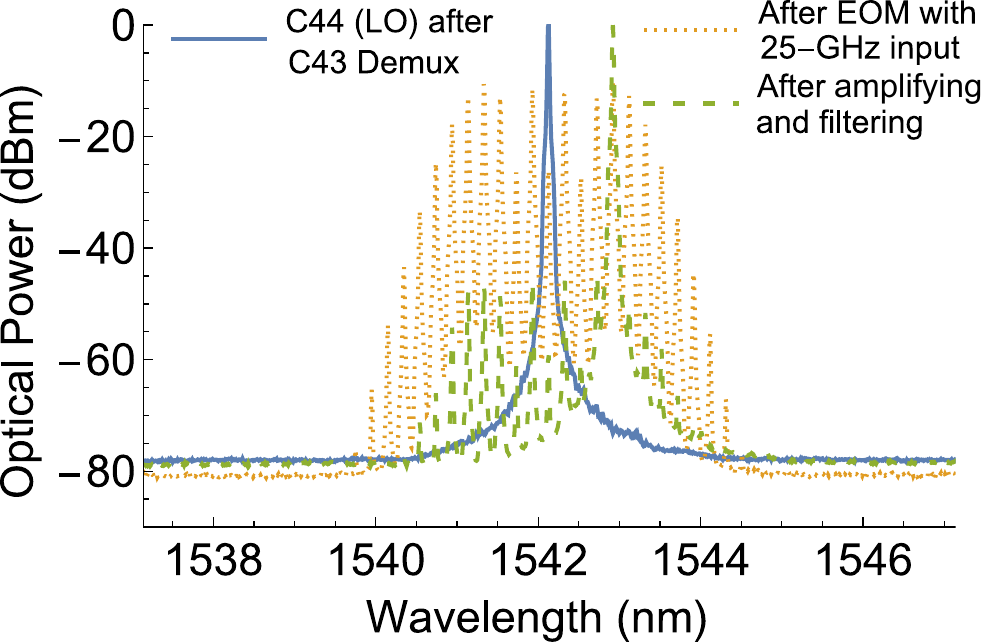}}
\caption{Local oscillator wavelength conversion. Spectra of local oscillator conversion from C44 to C43 using an electro-optic phase modulator to create optical sidebands which are then amplified and filtered. Spectra measured using a resolution bandwidth of 0.02~nm.}
\label{fig:sidebandsOSA} %fig:sidebandsOSA -> fig:OEmethods(e)
\end{figure}

Two-mode squeezing is measured with joint homodyne detection. This detection needs a coherent local oscillator (LO) at each of the squeezing wavelengths (C45 and C43); to derive the LO, the leftover SHG pump laser at C44 is used.  The LO at C44 is converted to C43 and C45 by creating optical sidebands on C44 using 40-GHz electro-optic phase modulators (EOMs) driven by an amplified 25-GHz sinusoidal signal [see Fig.~\ref{fig:sidebandsOSA}]; the fourth sideband is 100-GHz away from C44 and aligns with the next nearest channel (C43 or C45); this sideband is maximized by adjusting the temporal phase modulation amplitude. The RF oscillators generating the 25-GHz signal are synchronized by a 10-MHz reference signal distributed using a RFoF system; without such synchronization, no squeezing is obtained. During testing, we found it necessary for there to be a nearly constant phase relationship between the LO on each side. This required the 25-GHz signals applied to the EOMs to be coherent and to be the same frequency within the linewidth of the LO ($<10$ Hz).  Sometimes when the 25-GHz RF oscillators are in same location, they are also coupled directly using a coaxial cable to share a 10-MHz reference; this allows comparison between several measurements with and without RFoF. More RFoF details are provided in \ref{app:RFoF}. After generation, the sidebands are optically amplified by erbium-doped fiber amplifiers (EDFAs) and filtered by 0.2-nm bandpass filters to isolate the fourth sideband on the correct side. These amplified and filtered fourth sidebands are used as the LOs in homodyne detection to mix with C43 or C45 squeezing, respectively. Additional homodyne detection system details are provided in \ref{app:HD}.

Homodyne detection signals are jointly analyzed in two ways---in real-time using a hybrid junction (HJ) and an electronic spectrum analyzer (ESA), or in post-processing using signals recorded by digital sampling oscilloscopes. The path lengths of the LO and squeezed light should always be matched well within the coherence length of the detected squeezed light for optimal coherence; for our laser, there is optimal coherence and minimum phase noise when they are matched within about 1~m of fiber, which they are in this demonstration. Additionally, when the squeezing path lengths are matched, two-mode squeezing can be directly analyzed after the homodyne detector using a HJ and ESA; thus, any added fiber in the channels also needs to be matched for this analysis method.  Initially, the 5-km spools were different lengths by over 100~m. The longer spool was shortened and had pigtails fusion spliced on it so the signal propagation time of the two spools match to less than 1~ns. 

The pathlengths of each LO and its corresponding squeezed light are well matched by design, thus, the path lengths of the jointly distributed two-mode squeezed state can be asymmetric and/or the homodyne detectors do not need to be in the same location when using the post-processing method.   This  analysis method records the output of each homodyne detector separately using a fast low-noise digital sampling oscilloscope (500 Msamples/s for 4 ms at a time). Oscilloscope acquisition is triggered on a single 20-ns wide 2-V pulse; the pulse is applied to the LO with an arbitrary waveform generator (AWG%,Tektronix AWG710
) and 10-GHz amplitude modulator. The trigger signal is then detected by one of the single-detector monitor outputs in each homodyne detector. The trigger pulse is in the middle of the dataset so the center $10^5$ samples out of $2\times10^6$ samples are discarded. The rest of the data is then post-processed and converted to quadrature samples.  To coordinate measurements in separate locations with the oscilloscopes, a previously developed classical control plane~\cite{PRXQuantum.2.040304} is used. A central computer over TCP controls arming of the oscilloscope triggers, an AWG to create a trigger pulse, and the recording of the oscilloscopes acquired samples.

The post-processing averages every four oscilloscope samples to provide individual voltage points $V_{\text{avg}}[k]$. Then these points are normalized so that a vacuum input has mean zero ($\braket{q}_\text{vac}=0$) and the variance is 1/2 ($\Delta q^2_\text{vac} = 1/2$) to become quadrature values. Thus, a quadrature value is given by 
\begin{equation}
    q[k] = (V_{\text{avg}}[k]-SN_{\text{m}})\sqrt{\frac{1/2}{SN_{\text{var}}}}
\end{equation}
for all $k\in\{1,2,...,K\}$, where $K=475000$, $SN_{\text{m}}$ and $SN_{\text{var}}$ are the shot-noise mean and variance, respectively. 

We then combine the quadratures to analyze the joint quadrature signals ($q_1+q_2$ and $q_1-q_2$) which are oppositely correlated, i.e., one is squeezed while the other is anti-squeezed. Two-mode squeezing covariance theory is detailed in \ref{app:tmsvtheory}. In contrast to separately calculating the shot noise levels for each side and converting each detection to normalized quadrature values, when using the ESA and HJ, the measured shot-noise levels are equalized on the ESA (bypassing the HJ) by adjusting the EDFA currents and it is verified that the shot noise level with the HJ is the same. From the post-processed joint quadratures, a rolling variance is calculated for visual analysis. To account for any relative delays between $q_1$ and $q_2$, the delay (in units of quadrature samples) between $q_1$ and $q_2$ is optimized via direct search: the rolling variance is calculated for several different delays; then, the optimal delay is chosen as the delay which has the largest average absolute visibility between $q_1+q_2$ and $q_1-q_2$. In practice, the optimal delays are found to be non-zero due to slightly different triggering times because: the bandwidth of the detector's monitor output is limited (about 1 MHz); the trigger pulses are not necessarily equal heights; and the oscilloscopes trigger on the trailing/rising edge of the negative trigger pulses. 

When doing real-time analysis with the ESA, the relative phases between the squeezed light and the LO's are allowed to drift naturally due to the environment. Those oscillations are usually on the order of 1~Hz; this drift speed allows for recording several oscillations during the 4-s sweep time. When using the sampling oscilloscopes, the acquisition time is 4~ms so the phase of the C45 LO is driven using a piezo fiber phase shifter and 125-Hz triangle wave. 

\section{Results}
\begin{figure}%[t!]
\centerline{\includegraphics[width=1\columnwidth]{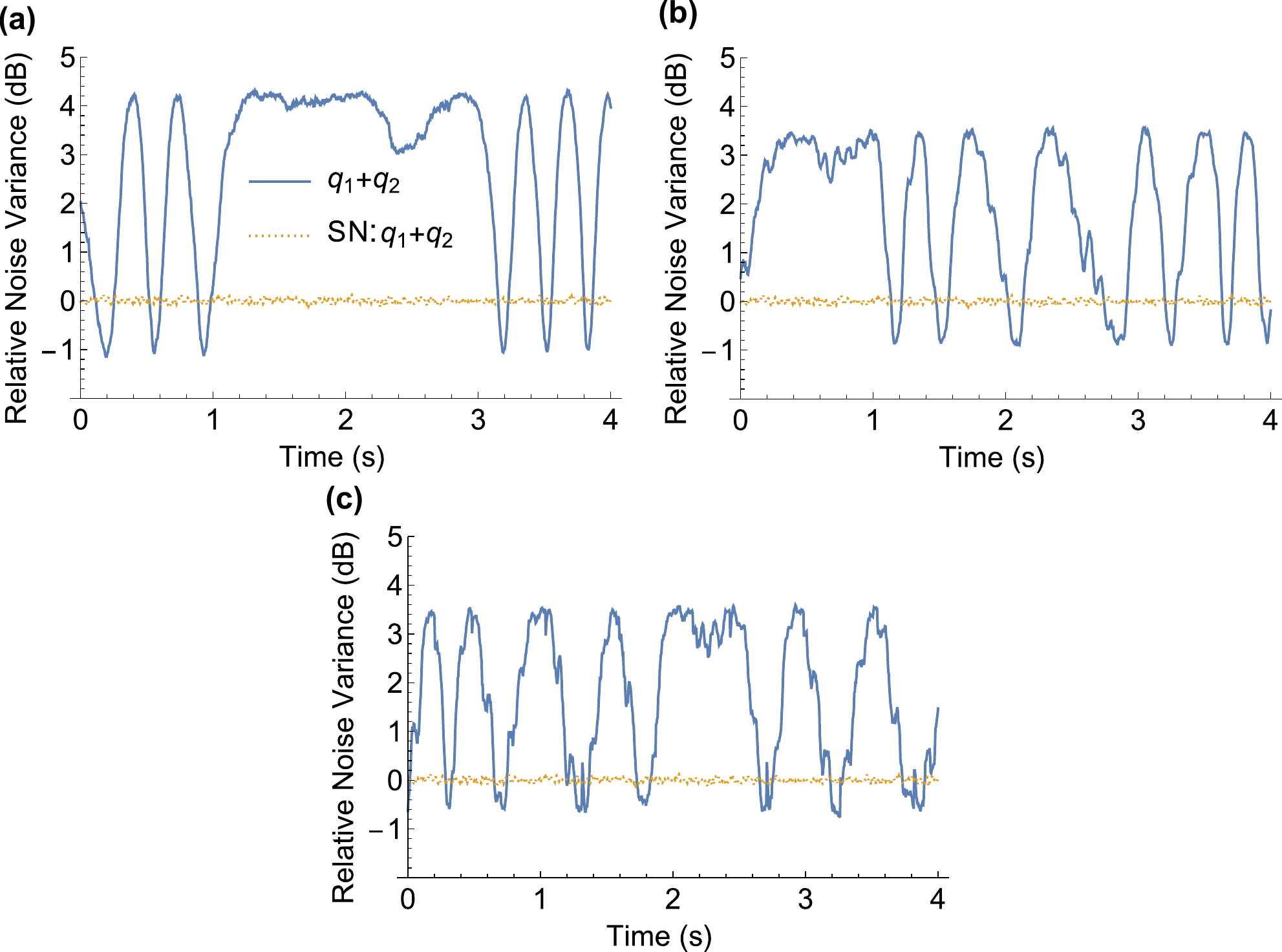}}
\caption{ESA-analyzed squeezing. (a) Reference data using 0-km channel length, no coexistence with conventional networking, and directly coupled RF oscillators. (b) 5-km channel length, coexistence with conventional networking, and directly coupled RF oscillators. (c) 5-km channel length, coexistence with conventional networking, and RFoF synchronized RF oscillators. All panels used zero-span ESA sweep centered at 8~MHz. Resolution bandwidth is 510~kHz. Video bandwidth is 20~Hz. Electronics noise is 15-dB below shot noise.}
\label{fig:ESAdata}
\end{figure}

Using an array of configurations---including different channel lengths, with(out) coexistence with other signals, different RF-oscillator frequency references, and different analysis methods---the performance of distributed squeezing is measured and compared. The reference configuration consists of four conditions: no added fiber (0-km channel length); no coexistence with conventional networking (but still multiplexed with the LO and RFoF signals); directly coupled RF oscillators instead of using RFoF; and use of the HJ and ESA for real-time analysis. In the reference configuration, average (anti-)squeezing (averaged maxima and minima from five sweeps) is measured to be -1.16$\pm0.025$~dB (4.29$\pm0.03$~dB); for a representative sweep, see Fig.~\ref{fig:ESAdata}(a). This configuration has a total loss of 5.09$\pm0.1$~dB (5.89$\pm0.1$~dB) for C43 (C45) squeezed light; using these losses and $r=0.986$, the (anti-)squeezing is estimated to be  -1.19$\pm0.03$~dB (4.40$\pm0.07$~dB). $r=0.986$ is the best-fit value, with coefficient of determination R$^2$=0.9995, from (anti-)squeezing versus power measurements; details are provided in \ref{app:powvssq}.

The total loss to the squeezed light has many contributing factors: imperfect waveguide-to-fiber coupling (2.3~dB), waveguide propagation loss of about 0.2~dB/cm for 2.5-cm waveguide assuming pairs were on average created in the middle (0.25~dB), imperfect detection efficiency (0.5~dB), effective loss due to electronics noise~\cite{Appel2007} being 15-dB below shot noise (0.14~dB), imperfect homodyne beamsplitter transmission (C43 0.4~dB and C45 0.5~dB), and imperfect transmission through optical system between waveguide and homodyne detection (C43 1.5~dB and C45 2.3~dB). As indicated by the 15-dB ratio of shot noise to electronics noise, the measurement system is capable of faithfully measuring much higher levels of squeezing. With improved transmission, measurement of substantially more squeezing is possible; the squeezing being generated in the waveguide, using $r=0.986$ (corresponding to about 0.7~W pump power), is inferred to be 8.5~dB. Transmission improvements are possible, which we discuss further below.

%\begin{figure}%[t!]
%\centerline{\includegraphics[width=1\textwidth]{OElabresults}}
%\caption{(a) ESA-analyzed reference data using 0-km channel length, no coexistence with conventional networking, and directly coupled RF oscillators. (b) ESA-analyzed, 5-km channel length, coexistence with conventional networking, and directly coupled RF oscillators. (c) ESA-analyzed, 5-km channel length, coexistence with conventional networking, and RFoF synchronized RF oscillators. (a)-(c) panels used zero-span ESA sweep centered at 8~MHz; resolution bandwidth is 510~kHz; video bandwidth is 20~Hz; electronics noise is 15-dB below shot noise. Separate-oscilloscope squeezing. (d) Separate-oscilloscope reference data using 0-km channel length, no coexistence with conventional networking, and directly coupled RF oscillators. (e) Separate-oscilloscope, 5-km channel length, coexistence with conventional networking, and directly coupled RF oscillators. (f) Separate-oscilloscope, 5-km channel length, coexistence with conventional networking, and RFoF synchronized RF oscillators. (d)-(f) panels used homodyne signals captured by separate oscilloscopes and post-processed to calculate the rolling variance (using 10k quadrature samples, corresponding to 0.08~ms). (d)-(f) panels show downsampled datasets, showing 250-evenly-spaced points out of 444960 quadrature samples, for visual purposes; electronics noise is 13-dB below shot noise.}
%\label{fig:OElabresults}
%\end{figure}

\begin{figure}%
\centerline{\includegraphics[width=0.5\textwidth]{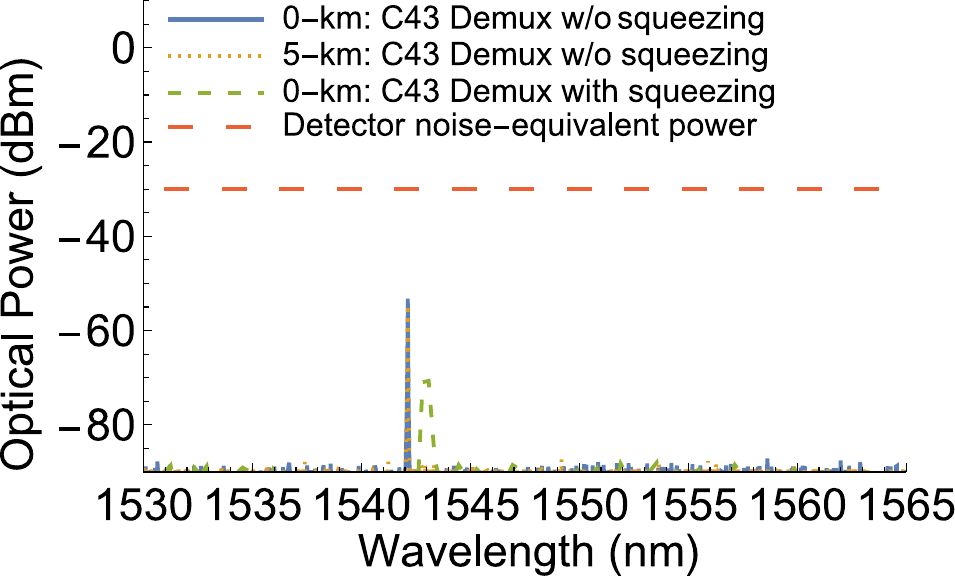}}
\caption{Demultiplexed detector input spectra. Spectral measurement of the signal going into the C43 homodyne detector. We measured the spectra for 0-km and 5-km with the squeezed-light blocked and 0-km with the squeezed light. We also include the measured noise-equivalent power of the balanced detector at C43 using the ESA. Spectra measured with an optical spectrum analyzer, using a resolution bandwidth of 0.02~nm.}
\label{fig:Demuxspec} %fig:Demuxspec -> fig:OEmethods(f)
\end{figure}

Moreover, the homodyne measurement of squeezed light could degrade from multiplexing with other signals in three main ways: (1) Light at the same wavelength as the LO would mix with the LO in the homodyne detection (at C43 or C45 after sideband conversion) and compete with the squeezing that is mixed with the LO; (2) Light at other wavelengths besides the LO could generate light at the LO wavelength, e.g., via four-wave mixing or Raman scattering, and mix with the LO in homodyne detection; (3) Light, at other wavelengths besides the LO, could contribute its own shot noise upon detection if it is greater than the noise equivalent power (NEP) of the balanced detector.   Fig.~\ref{fig:Demuxspec} displays the spectral measurement of the signal going into the C43 homodyne detector (i.e., after demultiplexing). We measured the spectra for channel lengths of 0-km and 5-km with the squeezed-light blocked and channel length of 0-km with the squeezed light unblocked. We also include the measured noise-equivalent power of the balanced detector at C43 using the ESA.
(1) and (2) are excluded because Fig.~\ref{fig:Demuxspec} shows no appreciable amount of light at the squeezing wavelength (C43 in this case) from the multiplexed signals---even when we add 5-km of fiber to the channel. (3) is also excluded because Fig.~\ref{fig:Demuxspec} also shows there is no light at other wavelengths above the noise-equivalent power of the balanced detector, which is about -30 dBm (as measured before the homodyne beamsplitter), i.e., the detector is much less sensitive to light that is not at the wavelength of the LO. In fact, with the noise floor of the OSA being -90 dBm at 0.02-nm resolution, any leakage from the conventional networking signal at H35 and RFoF at 1310~nm (separate measurement) is below the noise floor due to sufficiently high filter isolation. The leakage seen is from the C44 LO (before wavelength conversion); the leakage is well below the NEP. Thus, there is no expectation of degraded squeezing due to the multiplexed signals, beyond the added insertion loss for the act of (de)multiplexing.

Conventional networking signals coexisting with squeezed light and using 5-km fiber spools for Channel \#1 and \#2 is what distinguishes the next experimental configuration from the reference. In this configuration, averaged (anti-)squeezing is measured to be -0.96$\pm0.04$~dB (3.57$\pm0.08$~dB) [Fig.~\ref{fig:ESAdata}(b)]. The estimated expected (anti-)squeezing is -0.88$\pm0.02$~dB (3.7$\pm0.06$~dB); the estimate uses $r=0.986$ again and adds 1.1(1.2)$\pm0.1$dB loss to Channel 1(2) from the 5-km spools.  Furthermore, the  average measured (anti-)squeezing is -0.972$\pm0.024$~dB (3.62$\pm0.01$~dB) without having conventional networking signals coexisting through the 5-km fiber; thus, in agreement with the analysis of Fig.~\ref{fig:Demuxspec}, the coexisting conventional networking signals cause no statistically significant degradation due to the squeezing. Moreover, the mean conventional single-flow TCP throughput between the hosts during the various squeezed-light tests was in 9.746-9.750~Gbps range, with standard deviation in the range 0.012-0.014. With no squeezed light and 0-km~(5-km) added channel length, the mean throughput was 9.75$\pm0.01$~Gbps~(9.73$\pm0.01$~Gbps). Consequently, the TCP throughput is not statistically affected by the squeezed light. %}

In addition to coexistence and 5-km spools, the experimental configuration of Fig.~\ref{fig:ESAdata}(c) also uses the RFoF system to distribute the 10-MHz reference instead of the direct coupling used in Fig.~\ref{fig:ESAdata}(a)-(b). In Fig.~\ref{fig:ESAdata}(c), the RFoF system is shown to enable detection of squeezing, albeit with somewhat decreased visibility and increased phase noise. As such, the measured average squeezing over 5 sweeps is reduced to -0.722$\pm0.03$~dB; while the measured average anti-squeezing (3.61$\pm0.03$~dB) is relatively unchanged, since the measured squeezing is more sensitive to phase noise. This reduction in squeezing is also seen with no added fiber; so it is not due to the 5-km spools, though they do appear to add a small amount of extra phase noise themselves. The low-frequency (1-50~Hz offset) phase noise, measured using a 1-Hz resolution bandwidth, is nearly identical between the RFoF received signal, the transmitter, and the RF-oscillator internal reference; but, there is a significant difference in the total harmonic distortion (THD) of the 10-MHz references, which is most likely the cause of the difference in squeezing. Reduced squeezing is measured with 10-MHz references with increased THD, i.e., RFoF receivers (-19~dBc), compared to squeezing measured using 10-MHz references with lower THD, e.g., RFoF transmitter (-35~dBc) or RF-oscillator internal reference (-43~dBc). Further 10-MHz reference spectral analysis details are provided in \ref{app:RFoF}.  This analysis shows the 10-MHz reference provided by the RFoF system could be improved by incorporating a low-insertion loss narrowband 10-MHz filter with high-rejection tailored to this system to achieve no degradation in squeezing; though readily available, conventional RC low-pass filters do not provide enough harmonic rejection with low-enough insertion loss at 10-MHz to be used effectively in our system. Ultimately, the RFoF subsystem is only part of the creation of a phase-coherent LO for joint homodyne detection; so although there is a reduction of measured squeezing, it is not an indication of physically reduced squeezing. In fact, a shorter time-scale may help to measure the squeezing still present.

\begin{figure}%[t!]
\centerline{\includegraphics[width=1\textwidth]{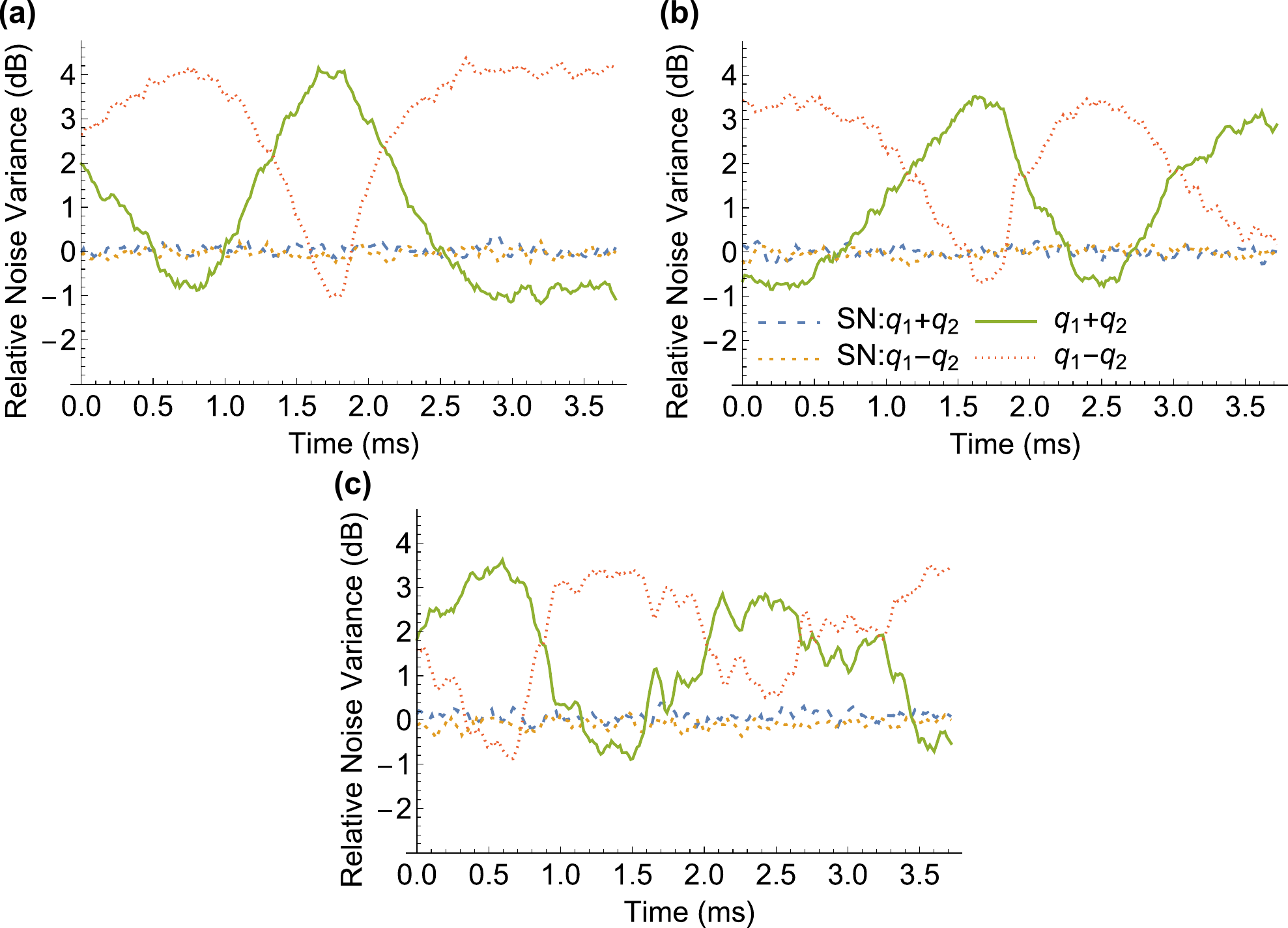}}
\caption{Separate-oscilloscope squeezing. (a) Reference data using 0-km channel length, no coexistence with conventional networking, and directly coupled RF oscillators. (b) 5-km channel length, coexistence with conventional networking, and directly coupled RF oscillators. (c) 5-km channel length, coexistence with conventional networking, and RFoF synchronized RF oscillators. All panels used homodyne signals captured by separate oscilloscopes and post-processed to calculate the rolling variance (using 10k quadrature samples, corresponding to 0.08~ms). Plot shows downsampled dataset, showing 250-evenly-spaced points out of 444960 quadrature samples, for visual purposes. Electronics noise is 13-dB below shot noise.}
\label{fig:oscdata} %-> fig:OElabresults (d-f)
\end{figure}

Using the same configurations used in Fig.~\ref{fig:ESAdata}, except the homodyne detection signals are recorded by the digital sampling oscilloscopes and post-processed, Fig.~\ref{fig:oscdata} displays noise variance sweeps.  Squeezing is clearly present in all configurations using post-processing of joint detection of separate homodyne systems; though, the timescale is much shorter due to the limited internal oscilloscope sample memory. This shorter timescale enables viewing of transient higher-frequency phase noise of the 25-GHz signal which appear to occasionally degrade the visibility as in Fig.~\ref{fig:oscdata}(c) and to a smaller degree on the left side of Fig.~\ref{fig:oscdata}(a). Since transients are seen without the RFoF system [Fig.~\ref{fig:oscdata}(a)], it is possible the 2.5:1 voltage-standing-wave ratio of the power amplifier before the EOM is causing added phase noise, with likely improvement using a different amplifier. Moreover, this timescale could enable measuring squeezing with only transitory degradation by the RFoF system; in fact, this is the case. In Fig.~\ref{fig:oscdata}(c), for the 5-km channel using coexistence and RFoF, we measure (anti-)squeezing of -0.9$\pm0.1$~dB (3.6$\pm0.1$~dB). The estimated expected (anti-)squeezing is -0.86$\pm0.02$~dB (3.7$\pm0.06$~dB); the expected squeezing is smaller by about 0.02~dB because the electronics noise of the oscilloscope is somewhat higher than the ESA.  For post-processed data, the error bar is the standard deviation of SN:$q_1\pm q_2$. The squeezing is clearly detectable with this method; it matches well with expectations and can even measure higher transient squeezing than the ESA. This shows squeezing can be measured using two completely separate homodyne detection systems with the joint squeezing (e.g., $q_1+q_2$) recovered in post-processing---an important capability for CV quantum networking.

%250 points were chosen to not cut out any major features but to simultaneously show visual differences between lines, i.e., dashing.

\begin{figure}%[t!]
\centerline{\includegraphics[width=1\textwidth]{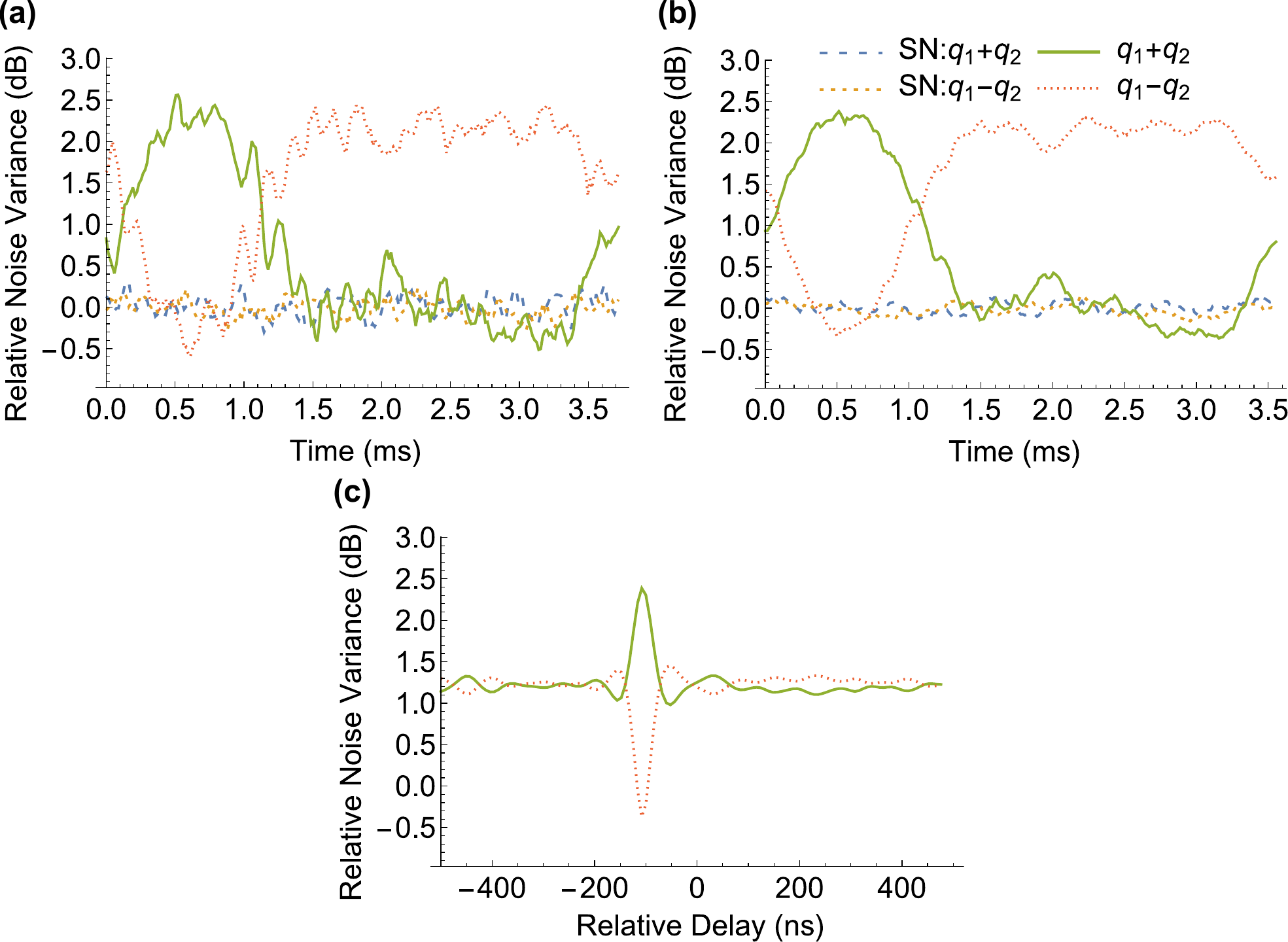}}
\caption{Deployed-fiber squeezing. (a) Rolling variance with 10k quadrature samples. (b) Rolling variance with 30k quadrature samples. (c) 30-ksamples variance versus relative delay between $q_1$ and $q_2$ at 0.51~ms [in Fig.~\ref{fig:deployeddata}(a)]. All panels used homodyne signals captured by separate oscilloscopes and post-processed to calculate the rolling variance. (a) and (b) show downsampled dataset, showing 250-evenly-spaced points out of 444960 quadrature samples, for visual purposes. Electronics noise is 13-dB below shot noise.}
\label{fig:deployeddata}
\end{figure}

Moreover, for Channel \#1 and \#2, deployed campus fiber is also used. This necessitated the use of the RFoF system to distribute a shared 10-MHz reference and analysis in post-processing after capture of the homodyne detection signals by the digital sampling oscilloscopes. In this configuration, the demultiplex filters and homodyne detection systems, including sideband generation, are located separately from each other and from the squeezed-light source. For a representative sweep with low transient higher-frequency phase noise and correspondingly high visibility [Fig.~\ref{fig:deployeddata}(a)], (anti-)squeezing is measured to be -0.5$\pm0.1$~dB (2.5$\pm0.1$~dB); this measurement is in agreement with the predicted (anti-)squeezing is -0.5~dB (2.6~dB) using only the total loss (C43: 9.77~dB and C45: 7.97~dB) and $r=0.986$. In Fig.~\ref{fig:deployeddata}(a) the variation of the shot noise traces is a significant amount of the measured squeezing. Fig.~\ref{fig:deployeddata}(b) shows even more statistically significant squeezing is present using the same data as Fig.~\ref{fig:deployeddata}(a) but with 30k samples included in the rolling variance instead of 10k. The squeezing shown in Fig.~\ref{fig:deployeddata}(b) is reduced because the increased samples in the rolling variance effectively averages over some of the oscillations seen in the $q_1+q_2$ and $q_1-q_2$ traces of Fig.~\ref{fig:deployeddata}(a) since 30k samples corresponds to 0.24~ms. The 30-ksamples variance at 0.51~ms [in Fig.~\ref{fig:deployeddata}(a)] is calculated versus relative delay between $q_1$ and $q_2$ [Fig.~\ref{fig:deployeddata}(c)]. This plot displays even more clearly the anti-correlation between $q_1+q_2$ and $q_1-q_2$. Moreover,  the squeezing is present only for a narrow set of delays. The squeezing coherence time can be estimated from the width of the dip and peak (40~ns full-width at half-maximum). The coherence time is expected to be shorter than the inverse of our detected squeezing bandwidth, which is about 12~MHz, limited by our homodyne detector bandwidth. Thus, we demonstrate distribution and measurement of two-mode squeezing multiplexed with multiple classical signals on deployed fiber. Within statistical variation, our measurements in Fig.~\ref{fig:deployeddata} agree with our estimate using only channel loss and $r$, i.e., no excess noise is observed.

\section{Discussion}
%Discuss improvements
In this work, two-mode squeezing is measured using joint homodyne detection located in separate buildings after distributing the squeezed modes through deployed fiber and multiplexing them with multiple classical signals, including the LO in a wavelength channel adjacent to the squeezing. The insertion loss of the multiplexing and demultiplexing is less than 2~dB using off-the-shelf filtering; but, the insertion loss could conceivably be improved with higher-performance filters and using a C-band RFoF transmitter. The latter allows all DWDM filters to be used so the CWDM filters can be removed from the squeezed-light paths, improving the squeezed-light transmission. The 2.3-dB waveguide-to-fiber coupling loss could be improved as significantly lower coupling has been demonstrated in~\cite{9044771}. Intrinsic loss of new variable beamsplitters is less then 0.1~dB; but, over months, the index-matching fluid between the polished beamsplitter fibers dries, increasing the insertion loss. The original insertion loss is not fully recovered with added index-matching fluid. Detection efficiency can be improved by using photodiodes with about 99\% quantum efficiency in the C-band~\cite{Mehmet:11}. With further optimization, the loss could realistically be reduced  by several dB. On the detection side, the phase noise of the RF signal from RF components and from the RFoF system 10-MHz reference can likely be improved using available technology.

In other respects, the local oscillator is not shot-noise limited in our detection bandwidth; this requires a sufficiently high common-mode rejection ratio (CMRR) of our homodyne detector for the balanced detection to cancel out the technical noise, so shot-noise is measured. Lower technical noise would enable a lower CMRR to still measure shot noise. Our homodyne detectors had sufficient CMRR (about 85~dB) but the coupling ratio of the variable beamsplitters was prone to drift at times based on environmental conditions; the VBSs are sometimes stable for hours, sometimes seconds. The CMRR stability could be increased with active stabilization. The ESA measurements are taken at 8~MHz where there is less technical noise; whereas, the oscilloscope measurements use the full bandwidth of the homodyne detector. Filtering, of the homodyne detection signals before the oscilloscope, to narrow the recorded bandwidth at a frequency range with low technical noise ( near 8~MHz) would also reduce the CMRR requirements. Nonetheless, we were careful during data acquisition to make sure our data was taken when the CMRR was sufficiently high and stable so the shot noise was not significantly affected; we also used conservative measurements of the shot noise as references for squeezing.

%Discuss future directions and wrap up
We demonstrated two-mode squeezing between a single mode pair, but the SPDC bandwidth is very wide [Fig.~\ref{fig:SPDC}]. Over 100~nm of this bandwidth can contain significant squeezing~\cite{doi:10.1063/5.0063118}. Thus, this device emits a large number of squeezed mode pairs; over 50 usable-pairs (assuming 100-GHz multiplexing), over 200 usable-pairs (assuming 25-GHz multiplexing), or even more with narrower mode spacing, ultimately limited by the LO linewidth. Our scheme to create distributed LO's for each mode is scalable because the original LO can be split and amplified after transmission. Ultimately, our method is only limited by the sidebands able to be generated. With high-power high-bandwidth phase modulators and RF oscillators, using a single EOM could possibly reach over 10~nm but not much more because of the Bessel function response limiting the power in the sidebands. With multiple EOMs, an optical-frequency-comb LO can be generated~\cite{Yang2021} and tailored to have a flat top over the entire C-band~\cite{Wu:13}. % For mode pairs at larger offsets, a fiber-optic parametric amplifier (FOPA) based on four-wave mixing (FWM) can be employed to create sidebands at much larger spacings, dependent on the FWM pump-signal spacing~\cite{Oda:07}. Care would need to be taken to ensure the phase coherence of the LO is transferred to the sideband of interest since some FOPA configurations are phase insensitive~\cite{Andrekson:20}. 
Alternatively, resonant electro-optic frequency combs using microring resonators~\cite{Zhang2019,Hu2022,Chang2022} may be a more promising method to generate sidebands far away from the original LO. Phase coherence of the LO would need to be preserved in whatever configuration used. Also, each LO of a squeezed-mode needs to have complimentary spacing from the degenerate squeezing wavelength to jointly measure the signal and idler from the same squeezed mode.

%conclusion
Here, we conclusively demonstrate the capability to distribute two-mode squeezing and use triggered joint homodyne detection with the source and each homodyne detection system located in separate buildings that are connected by fiber that has conventional network traffic on it as well. Thus, this work demonstrates the feasibility of two-mode squeezing distribution on fiber shared with conventional networking signals and the capability to use post-processed joint homodyne detection on physically separate homodyne detectors; both important capabilities necessary for future practical protocol demonstrations using lit fiber between physically separate locations.  Furthermore, any realistic CV quantum network using multi-mode squeezing will need the squeezed modes to propagate away from the source to various physically separated network nodes. For instance, this work enables development of truly distributed quantum sensing and networking based on two-mode squeezing which will benefit from removing the need for co-located joint detection.

\appendix
%\renewcommand\thesection{\alpha{section}}
%\newcounter{appcntr}
%\addtocounter{appcntr}{0}

%\renewcommand{\appendixsec}[1]{\section{#1}}
\renewcommand{\thesection}{Appendix \Alph{section}}

\section{Laser and waveguide subsystem} \label{app:lwsubsys}
To produce broadband squeezing, a fiber-coupled continuous-wave 771-nm second-harmonic generation (SHG) (NKT Photonics Harmonik 1W) pumps a fiber-coupled 2.5-cm type-0 periodically poled lithium-niobate (PPLN) ridge waveguide (AdvR RSH-T0771-P15P85AB0); this allows for the leftover SHG pump laser at 1542 nm (NKT Photonics Koheras Adjustik X15) to be coherently used as the local oscillator (LO). The 771-nm pump derived from second-harmonic generation (SHG) is stable in power to about 1\%. We filter the light to only allow 771~nm, and not other harmonics or fundamental, light through then coupled it into single-mode fiber. The leftover 1542-nm laser is also filtered to reject all harmonics after the SHG module and attenuated using a waveplate and polarizing beamsplitter to allow for variable attenuation. The 1542-nm laser is then coupled into polarization-maintaining single-mode fiber and sent through a 10-GHz amplitude modulator (EOSpace). We need significant attenuation of the leftover 1542-nm light; so the output is very sensitive to small polarization fluctuations. Thus, we pick off some power after the amplitude modulator using a 95/5 fiber coupler and direct it to a power meter (Thorlabs S120C and PM400). The reading on the power meter is collected by a python program running a control algorithm~\cite{simplepid}. The error signal is used to change the voltage source (Thorlabs MDT693B) connected to the DC bias of the amplitude modulator. The amplitude-modulator output power is stabilized to less than 0.1\% fluctuation which is important because it is used as the local oscillator in the rest of the experiment.

The waveguide temperature is temperature controlled (Thorlabs TC200) to about 27$^{\circ}$C, chosen to optimize the SHG in the waveguide using the LO as a pump. When using the 771-nm pump to produce spontaneous parametric downconversion (SPDC) in the waveguide, we filter the 771-nm pump from the SPDC using a free-space long-pass filter (Semrock BLP01-830R-25) in a fiber-coupled U-bench (Thorlabs FBC-1550-APC). This small free-space area also allowed a small anodized aluminum beam block to be placed in the squeezed-light path after the filter so that the shot-noise could be measured when needed, without need to block the squeezed-light pump or disconnect any fibers.

Due to the intensity of the signals, we found it necessary to record the spectra of C44 and H35 in Fig.~\ref{fig:CHspec} using a 0.1-nm full-width at half-maximum 30-dB notch filter (Finisar Waveshaper 1000A) inline before the spectrometer (Yokogawa AQ6370B) because without such a notch filter light from the main spectral peak scatters into the spectrometer's detector while the grating is sweeping through other wavelengths which creates an artificial spectrally broadband background, which can be seen in the blue trace of Fig.~\ref{fig:sidebandsOSA} where a notch filter was not used because the focus of the figure is the peak, not the pedestal. We calibrate the notch filter by first recording the spectra of a broadband amplified spontaneous emission (ASE) source (Pritel FA-30-IO) filtered using a 40-nm bandpass filter (Pritel TFA 40 nm) that passes light in the C-band then we record the spectra of the broadband ASE after going through the notch filter. This allows us to calibrate the filter transfer function including the insertion loss.

\section{Conventional traffic and classical link description} \label{app:conventraff}

\begin{table}
\caption{Classical signal propagation losses. Refer to Fig.~\ref{fig:setup} for measurement location within rest of experiment. AM = amplitude modulator. APD = avalanche photodiode. EDFA = erbium-doped fiber amplifier. LO = local oscillator. RFoF = radio frequency over fiber. Rx = receiver. VBS = variable beamsplitter. XFP = 10-Gbps small-form-factor pluggable transceiver.}
    \centering
    \begin{tabular}{l l S[table-format=3.2]}%d{3.2}}
    %\hline
    \hline
    Signal& Measurement location & {Power (dBm)}  \\
    \hline
    LO & After AM & 13.3\\
    LO & Into C45 quantum channel & 6.11\\
    LO & Into C43 quantum channel & 6.7\\
    LO & Before C45 EDFA & 0\\
    LO & Before C43 EDFA & 1.1\\
    LO & Before C45 VBS & 3.4\\
    LO & Before C43 VBS & 3.2\\
    \hline
    RFoF & Out of Tx & 4.14\\
    RFoF & Into C45 quantum channel & -12.76\\
    RFoF & Into C43 quantum channel & -10.1\\
    RFoF & Into C45 APD & -14.2\\
    RFoF & Into C43 APD & -10.9\\
    \hline
    XFP & Host \#1 Tx after all-optical switch & 0.46\\
    XFP & Host \#1 in squeezing source lab & -3.4\\
    XFP & Host \#1 into C43 quantum channel & -5.2\\
    XFP & Host \#1 after C43 demultiplexing & -7\\
    XFP & Host \#1 all-optical switch Rx & -11.85\\
    XFP & Host \#2 Tx after all-optical switch & 1.31\\
    XFP & Host \#2 in squeezing source lab & -4.4\\
    XFP & Host \#2 into C45 quantum channel & -6.2\\
    XFP & Host \#2 after C45 demultiplexing & -9.35\\
    XFP & Host \#2 all-optical switch Rx & -14.3\\
    \hline
   % \hline
    \end{tabular}
    
    \label{tab:loss}
\end{table}

The classical signals are generated between two Supermicro Linux servers equipped with 10GigE network interface cards (NICs) with 850-nm transceivers that are connected to two switches (Cisco 3064) over multi-mode fiber connections, respectively.  The classical  traffic is generated between two 10GigE 1550nm XFPs plugged into respective switches, and these XFPs are connected to an all-optical switch (Huber+Suhner Polatis)  which is connected to single-mode fiber to WDM and OADM devices. Each 850-nm multi-mode Ethernet server connection is cross-connected to an XFP on the corresponding Cisco switch. Thus, a conventional 10-Gbps network path is created between the two servers and it carries the conventional traffic on the single-mode fiber path between XFPs on the switches, which is used for coexistence testing with the squeezed light. TCP traffic is sent between the servers using the iperf3 tool that sends memory contents from server to client using the TCP protocol. This TCP traffic flow results in Ethernet packets being exchanged between the XFPs over the shared single-mode fiber, which constitutes the classical network traffic. All the conventional networking equipment is located in a separate conventional networking lab and is connected via fiber to the labs where the squeezing is generated and measured. The classical signal (LO, XFP, and RFoF) propagation losses are characterized and shown in Table~\ref{tab:loss}; this can be used to calculate insertion losses for various components or groups of components by referring to Fig.~\ref{fig:setup} for context of the measurement location within the rest of the experiment. In addition to these losses, the all-optical switch has about 0.5~dB insertion loss and 8-dB  of attenuators were added before each XFP receiver to avoid receiver saturation.

\section{Radio-frequency transfer over fiber}\label{app:RFoF}
A radio-frequency over fiber transmitter (Tx, RFoptic RFoF2T3FT-PA-11), with the input connected to a 13-dBm 10-MHz reference signal (SRS DS345), is used to distribute a shared 10-MHz reference. The 10-MHz signal is detected by modified amplified avalanche photodiodes (RFoF Rx, Thorlabs PDB570C), the feedback resistor of the first-stage transimpedance amplifier was changed from 1-kOhm to 10-kOhm. The 1310-nm RFoF transmitter is sorted by the DWDM filters in the multiplex and demultiplex systems with high insertion loss since the filters are designed for C-band signals. Thus, it is important to have high input RF power and high receiver amplification so the output 10-MHz signal can be about 0~dBm which is required for the RF oscillators to lock onto the 10-MHz reference. During development, the amplifier noise floor, if high enough, was shown to contribute phase noise so avalanche photodiodes and low-noise amplifiers were chosen to minimize the added phase noise while still providing the amplification necessary so the 10-MHz reference could be locked onto by the RF oscillators.

\begin{figure*}%[t!]
\centerline{\includegraphics[width=1\textwidth]{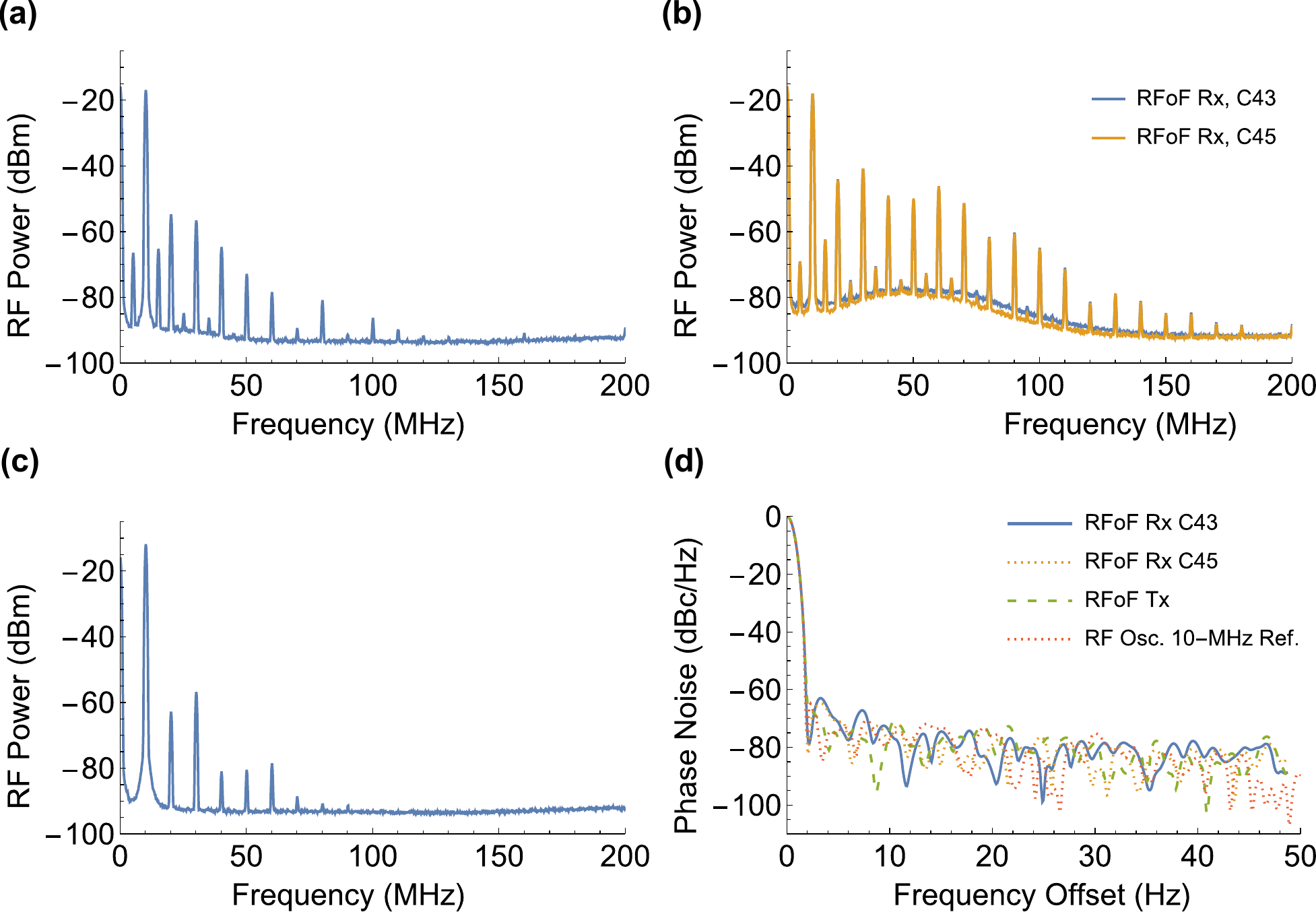}}
\caption{RFoF frequency-domain analysis. RF power versus frequency using ESA with 2-MHz resolution bandwidth and 1-kHz video bandwidth for (a) Input to RFoF transmitter after 30-dB attenuation, (b) RFoF receivers after 20-dB attenuation, and (c) RF-oscillator internal 10-MHz reference  after 20-dB attenuation. (d) Phase noise at low-frequency offsets using 1-Hz resolution bandwidth and 1-Hz video bandwidth.}
\label{fig:RFoFfd}
\end{figure*}

Fig.~\ref{fig:RFoFfd}(a)-(c) compares the spectral power of the RFoF transmitter, RFoF receivers, and the RF-oscillator internal 10-MHz reference (after an attenuator to avoid ESA saturation). The main 10-MHz peaks are identical within the resolution bandwidth of the ESA. This is shown even more clearly in Fig.~\ref{fig:RFoFfd}(d) where the phase noise for low-frequency offsets is plotted; in this graph, all signals' phase noise are less than the resolution bandwidth of the ESA (1-Hz for this measurement). But, in Fig.~\ref{fig:RFoFfd}(a)-(c), a difference is observed in the harmonics and spurs. This difference is quantified more precisely in Table~\ref{RFoFtable} using total harmonic distortion (THD) and spur-free dynamic range (SFDR). The RF-oscillator internal reference does have the lowest THD and highest SFDR, followed by the RFoF transmitter and then the receivers which have significantly degraded THD and SFDR. These extra harmonics and spurs in Fig.~\ref{fig:RFoFfd}(b) are likely due to clipping of the RF signal within the transmitter itself. The RFoF transmitter has a 1-dB compression point of -3~dBm which is much less than the 13~dBm found to be the optimal power to put into it. Testing determined it was better to have higher RF input power to the transmitter than to increase the gain at the receiver even though this does introduce some clipping to the RF signal. Although, the SRS DS345 did not produce the cleanest 10-MHz reference for the RFoF transmitter that was readily available during development, it did produce the cleanest reference that could also produce high enough input RF power. There is also a noticeable noise hump centered at about 50~MHz in Fig.~\ref{fig:RFoFfd}(b); this is due to the amplification after the receiver avalanche photodiodes. Avalanche photodiodes have internal low-noise amplification that linear photodiodes do not. Thus, using avalanche photodiodes produced higher RF signals with less amplification necessary by RF amplifiers. When testing other amplifiers that raised the noise floor by 10-40-dB after the linear-photodiode receiver, the visibility and phase noise of the squeezing was noticeably degraded, sometimes to the point that no squeezing is  measurable.

\begin{table}
\caption{Analysis of 10-MHz reference spectra. THD (Total Harmonic Distortion). SFDR (Spur-free dynamic range).}
\label{RFoFtable}
\centering
\begin{tabular}{l S[table-format=3.2] S[table-format=3.2]}
\hline
%\hline
Signal & {THD (dBc)} & {SFDR (dBc)}\\
\hline
RFoF Tx & -34.8 & 37.8\\
RFoF Rx C43 & -19.3 & 22.9\\
RFoF Rx C45 & -19.2 & 22.8\\
RF Osc. Int. Ref. & -43.3 & 45\\
\hline
%\hline
\end{tabular}
\end{table}

Additionally, the RF oscillators appear to have a significant amount of internal averaging of the 10-MHz reference which, in conjunction with the increased phase noise from the RFoF system, causes a reduction in measured squeezing. This conclusion was tested by retaking measurements with other RF oscillators (SignalCore SC5511A) which presumably have lower internal averaging of the 10-MHz reference but have higher phase noise themselves and this configuration does not show a reduction in squeezing when the RFoF output is used as their 10-MHz reference input. Ultimately though, we decided to use the initial RF oscillators due to their lower phase noise even though there was some reduction in the measured squeezing when using the RFoF system. These measurements show that the reduction in measured squeezing is not inherent to the RFoF system but also is due to the equipment using the 10-MHz reference.

Given the very comparable low-frequency phase noise, down to 1-Hz, and the significantly different THD and SFDR, it appears that narrow filtering of the 10-MHz reference would decrease the phase noise apparent in the squeezing data. Future work to improve the RFoF system includes incorporating a 10-MHz bandpass filter tailored to this system or replacing the transmitter with one that has a higher 1-dB compression point. The ideal filter would require fast roll-off for high rejection of the nearby spurs and harmonics as well as have very low insertion loss for the 10-MHz peak so there is still enough 10-MHz-reference power for the RF oscillator to lock onto it. %Conventional RC low-pass filters, even higher order ones, do not have low-enough insertion loss for 10-MHz with simultaneously high-enough rejection of nearby spurs and harmonics; development of a specialized high-performance filter is required.
Moreover, there are multiple types of RFoF transmitters, those which directly modulate the laser-diode current and those which use modulator(s) after the laser. Our directly modulated transmit laser did show increased phase noise before being warmed up; using a system with a modulator after the laser may enable lower total transmit phase noise. Another option which may have even lower phase noise is to use RFoF to directly distribute 25-GHz instead of a 10-MHz reference that needs to be converted to 25-GHz by separate oscillators. There are increased costs for higher-frequency RFoF systems but RF oscillators are comparable in cost or more expensive so there are tradeoffs to be considered. In our case, we already had the RF oscillators and the lower frequency RFoF system from other experiments so our current configuration was a natural first demonstration for us.

\section{Homodyne detection and analysis}\label{app:HD}
To convert the LO from C44 to C43 and C45, using 40-GHz phase modulators (EOSpace) driven by an amplified (Pasternack PE15A4021) 25-GHz sinusoidal signal (Agilent E8257D) sidebands were created; the sideband amplitudes following the Bessel functions of the first kind via Jacobi–Anger expansion~\cite{Abramowitz1988}. The fourth-sideband optical power is maximized by adjusting the temporal phase modulation amplitude $\varphi(t) = \pm \Theta \sin\Delta\omega t$ ($\Theta = 5.31$~rad). With the estimated EOM half-wave voltages of $V_\pi = 5.65$~V at 25~GHz, the total radio-frequency (RF) power required at each EOM is about 29.6~dBm. The sidebands are then optically amplified  using erbium-doped fiber amplifiers (Pritel C43: SCG-40 and C45: FA-30-IO) and a 0.2-nm bandpass filter (Finisar Waveshaper C43:4000A and C45:1000A) is applied to isolate the fourth sideband on the correct side.

Our homodyne detectors use a variable fiber beamsplitter (VBS, C43: Oz Optics VBS-22-1550-8/125-P-3A3A3A3A-1-1-TLT and C45: Newport F-CPL-1550-N-FA) and an amplified balanced photodiode (Thorlabs PDB425C). The measured quantum efficiency of the photodiodes in our detectors is about 89\%. Moreover, we modified the electronics of both of the PDB425C modules to optimize them for our application: (1) We AC-coupled the input of the amplifier after the current summing junction using a 100-pF ceramic capacitor and 6-kOhm resistor; (2) We replaced the first-stage amplifier (Analog Devices OPA847) with a lower current-noise amplifier (Analog devices OPA657); (3) we increased the first-stage gain by using a 100-kOhm feedback resistor; and (4) We removed the second-stage amplifier. The modifications significantly reduced the electronics noise and increased the common-mode rejection ratio (CMRR); but as a consequence, the detection bandwidth is reduced from DC-75~MHz to 250~kHz-15~MHz. The AC-coupling is critically important for filtering out low-frequency technical noise from the laser, 5-km fiber spool, and 125-Hz oscillations that were slightly mechanically coupled from the piezo fiber phase shifter (we also physically isolated the phase shifter using foam).

\begin{figure}%[t!]
\centerline{\includegraphics[width=0.5\textwidth]{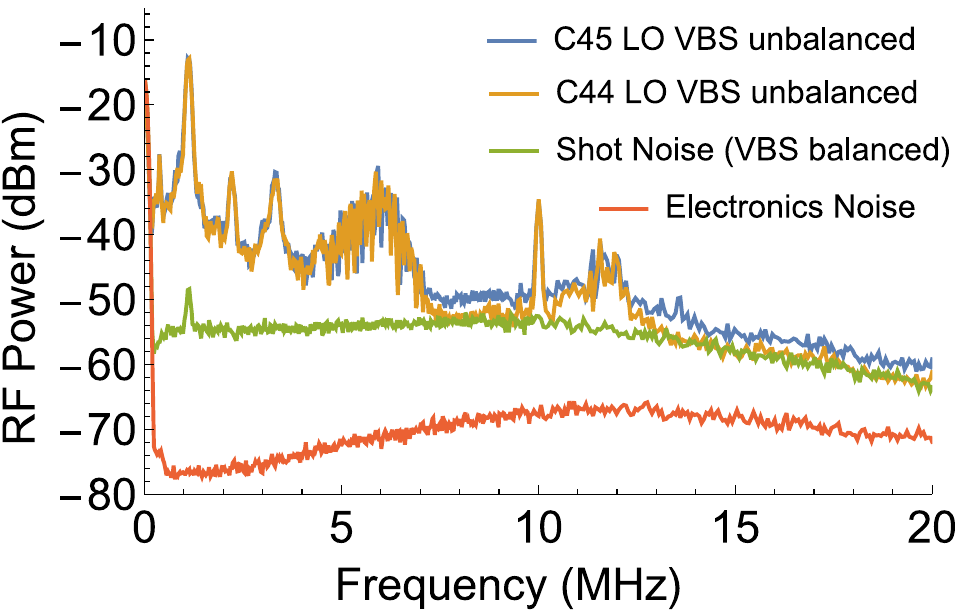}}
\caption{Frequency-domain noise measurements. Comparison of the technical noise of the LO before (C44) and after (C45) sideband conversion with each other, as well as with the shot noise (measured when the VBS is balanced) and the electronics noise.}
\label{fig:LOtn}
\end{figure}

When the VBS is slightly unbalanced (lower CMRR) technical noise of the LO is observed [see Fig.~\ref{fig:LOtn}]. The technical noise of the sideband-converted LO at C45 is nearly identical to the C44 LO before sideband conversion, except for places with low technical noise where there is an increased broadband background; This broadband background is likely due to ASE from the EDFAs used for sideband amplification or increased shot noise from more power onto one of the photodiodes. Fig.~\ref{fig:LOtn} shows that the sideband conversion does not significantly introduce technical noise. When the VBS balancing is optimized, the CMRR is high enough to cancel out the technical noise so that shot noise dominates, except for the highest peak; even then, for more optimized VBS splitting, the highest peak has been reduced below shot noise in other measurements.

\begin{figure}%[t!]
\centerline{\includegraphics[width=1\textwidth]{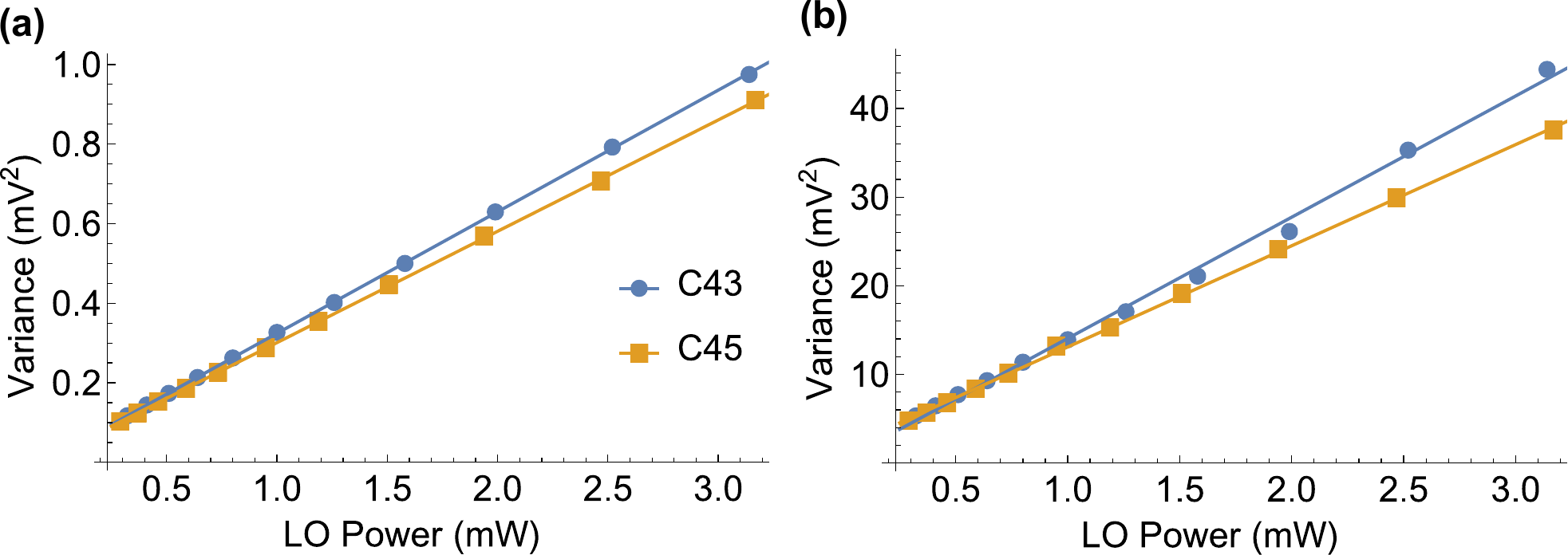}}
\caption{Shot-noise calibration. (a) Recorded using ESA centered at 8-MHz with 510-kHz resolution bandwidth and 20-Hz video bandwidth. Linear best-fit coefficient of determination R$^2$ is over 0.9999 for both. (b) Recorded by separate oscilloscopes and post-processed. Linear best-fit coefficient of determination R$^2$ is over 0.996 for both.}
\label{fig:SNcal}
\end{figure}

The shot noise of our homodyne detectors is calibrated for a range of relevant optical powers (changed by adding attenuation in the programmable 25-GHz filters) for both the ESA analysis at 8-MHz and post-processing the separate oscilloscope data [see Fig.~\ref{fig:SNcal}]. When using the ESA [Fig.~\ref{fig:SNcal}(a)], the RF power is measured, centered at 8 MHz with 510-kHz resolution bandwidth and 20-Hz video bandwidth, which is then converted to the variance of the voltage assuming a 50-Ohm load.  With the oscilloscopes, two million samples are recorded (spanning 4 ms at our 500-MHz sampling rate) for each output of the heterodyne detector. The center $10^5$ samples are discarded due to the trigger pulse. The samples are averaged in adjacent groups of four; the average values are saved as single quadrature points. From these quadrature samples, the mean, $SN_{\text{m}}$, and shot-noise variance, $SN_{\text{var}}$, are calculated for each detector. In Fig.~\ref{fig:SNcal}(b), slight drift in the common-mode rejection ratio (from the drift in the splitting ratio of the VBS) can be seen in the reduced linearity of the C43 shot-noise measurement compared to Fig.~\ref{fig:SNcal}(a) since the post-processed oscilloscope samples do not have the filtering of the ESA; this filtering removes some technical noise that is present sometimes when the VBS splitting ratio drifts to allow technical noise to exceed shot noise at certain frequencies. There is a slight difference between the slopes of the two homodyne detectors, likely due to not exactly matched amplifier gain; this difference does not appear to be significant given how the shot noise is calibrated for our squeezing measurements and how closely they match our expectations. Even still, equalizing the gains would likely somewhat improve the squeezing and anti-squeezing.

%Alex, write about passive VBS stability and how we tuned the VBS for maximum CMRR
Extensive care was taken to make small tuning adjustments to the splitting ratio of each VBS to attain the maximum CMRR for data acquisition. The primary challenge during the experiment arose from drift of the splitting ratio, i.e., reduction in the CMRR, which would occur on the order of seconds to hours after tuning. To help minimize the drift in the VBSs, each VBS and the input/output fibers are physically isolated from the environment by using cardboard boxes and foam packaging. The input/output fibers are looped and enclosed in cutouts in the foam. Holes were cut into the sides of the cardboard box to allow the fiber terminals to exit the packaging to connect with their respective incoming channels and photodiode. To improve stability, the fiber connectors leading to the VBSs are taped to the table on which the VBS's boxes sat; this is done to minimize the potential drift from vibrations as the splitting ratio would drift if the fibers are touched or moved. The non-polarization-maintaining-fiber-based VBS (Newport F-CPL-1550-N-FA) is often more stable in maintaining the maximum CMRR compared to the polarization-maintaining-fiber-based VBS (Oz Optics VBS-22-1550-8/125-P-3A3A3A3A-1-1-TLT). For future designs, using a mechanical-adjustment closed-loop-feedback system to balance the VBS may help avoid tuning past the max CMRR and stabilize to the max CMRR. Additionally, the device took time to settle at a new tuning location; when tuning the splitting ratio, after turning the knob a bit, a delay was observed before adjustments were observed in the splitting ratio (via CMRR measurements). There was also some time, on the order of seconds, before the splitting ratio would settle to a more stable value. Being able to characterize and predict this effect can improve the alignment process. After optimizing the VBS coupling ratio, the measured CMRR is about 85~dB or a little more, due in part to the AC-coupling of the detector which adds about 40~dB to the CMRR.

Joint analysis of the homodyne detection signals is done in two ways: in real-time using a hybrid junction (HJ, MA-COM HH-108) combined with an electronic spectrum analyzer (ESA, Agilent N9000A CXA Signal Analyzer) or post-processing using signals recorded by digital sampling oscilloscopes (C43: Agilent MSO-X 4104A and C45: Keysight MSOX4104A); post-processing is required when the squeezed-mode optical path lengths are not well matched, or the homodyne detectors are not in the same location.

For the post-processing method, the output of each homodyne detector is recorded separately on a fast low-noise digital sampling oscilloscope (500Msamples/s for 4 ms at a time). Then, acquisition is triggered on a single 20-ns wide 2-V pulse that is applied to the LO using an arbitrary waveform generator (Tektronix AWG710) and 10-GHz amplitude modulator (EOSpace) and is seen in one of the single-detector monitor outputs of the balanced detector which is connected to the sampling oscilloscope on its own channel. When using the sampling oscilloscopes for analysis, the acquisition time is 4~ms so the phase of the C45 LO is driven using a piezo fiber phase shifter (General Photonics) that is driven by a piezo driver (Thorlabs MDT694A) and 125-Hz 3-V amplitude 1.5-V offset triangle wave from the C45 sampling oscilloscope internal function generator.  Whereas, when the homodyne detection is analyzed in real-time with the ESA, the relative phases between the squeezed light and the LO's are allowed to drift naturally due to the environment. Those oscillations are usually on the order of 1~Hz which allows recording several oscillations during the 4-s sweep time.

\section{Two-mode squeezing covariance theory}\label{app:tmsvtheory}
To discuss two-mode squeezing and their covariances, the Gaussian-state formalism is used; we will start with a succinct introduction of this formalism and our notation. The convention of Ref.~\cite{RevModPhys.84.621} is followed, where Gaussian states are completely described by their displacement vector $\Bar{\textbf{x}}$ and their covariance matrix $\textbf{V}$. 
\begin{equation}
    \Bar{\textbf{x}}:=\text{Tr} (\Hat{\textbf{x}}\rho),
\end{equation}
where $\Hat{\textbf{x}}=(\Hat{q}_1,\Hat{p}_1,...,\Hat{q}_N,\Hat{p}_N)^T$ and the set $\{\Hat{q}_k,\Hat{p}_k\}_{k=1}^N$ are the quadrature field operators for $N$ modes. Thus, the covariance matrix is
\begin{equation}
    V_{ij}:=\frac{1}{2}\langle \{\Delta \Hat{x}_i, \Delta \Hat{x}_j \}\rangle,
\end{equation}
where $\Delta \Hat{x}_i:=\Hat{x}_i-\langle \Hat{x}_i \rangle$ and \{,\} is the anticommutator. Accordingly, a Gaussian unitary is described by these transformations:
\begin{equation}
    \Bar{\textbf{x}} \rightarrow \textbf{S}\Bar{\textbf{x}}+\textbf{d}\text{, } \textbf{V}\rightarrow \textbf{S}\textbf{V}\textbf{S}^T, \label{eq:transforms}
\end{equation}
where $\textbf{d}\in \mathbb{R}^{2N}$ and $\textbf{S}$ is a $2N\times 2N$ real matrix. To preserve the bosonic commutation relations (with $\hbar=2$), $\textbf{S}$ must be symplectic, i.e.,
\begin{equation}
    \textbf{S}\boldsymbol{\Omega}\textbf{S}^T =\boldsymbol{\Omega},
\end{equation}
where 
\begin{equation}
    \boldsymbol{\Omega}=\overset{N}{\underset{k=1}{\oplus}}\boldsymbol{\omega}=\begin{pmatrix}
    \boldsymbol{\omega} & &\\
    & \ddots &\\
    & & \boldsymbol{\omega}\\
    \end{pmatrix}\text{, }\boldsymbol{\omega}:=\begin{pmatrix}
   0&1\\
   -1&0\\
    \end{pmatrix},
\end{equation}
known as the symplectic form. In this convention, shot noise has a variance of 1.

Let us now consider four modes $\{a,b,c,d\}$. Modes $b$ and $c$ share a two-mode squeezed state of the form
\begin{equation}
    \textbf{S}_{\textbf{2}}(r):=\begin{pmatrix}
   \cosh{r}\textbf{ I}&\sinh{r}\textbf{ Z}\\
      \sinh{r}\textbf{ Z}&\cosh{r}\textbf{ I}\\
    \end{pmatrix},\label{eq:S2}
\end{equation}
where $\textbf{I}$ is the $2\times2$ identity matrix and $\textbf{Z}:=$ diag(1,-1). Whereas, modes $a$ and $b$ share a beamsplitter \textbf{B}({$T_b$}) to simulate imperfect transmission of mode $b$ and modes $c$ and $d$ share a beamsplitter \textbf{B}({$T_c$}) to simulate imperfect transmission of mode $c$. Here a beamsplitter between $N=2$ modes is modeled using the symplectic map
\begin{equation}
    \textbf{B}({T}):=\begin{pmatrix}
    \sqrt{{T}}\textbf{ I} & \sqrt{1-{T}}\textbf{ I}\\
    -\sqrt{1-{T}}\textbf{ I} &  \sqrt{{T}}\textbf{ I}\\
    \end{pmatrix}.\label{eq:BS}
\end{equation}
In this case, a balanced beamsplitter corresponds to ${T}=1/2$. In the full $N=4$ picture, where $\Hat{\textbf{x}}=(\Hat{q}_a,\Hat{p}_a,\Hat{q}_b,\Hat{p}_b,\Hat{q}_c,\Hat{p}_c,\Hat{q}_d,\Hat{p}_d)^T$, the two-mode squeezing of modes $b$ and $c$ is represented by the symplectic map
\begin{equation}
    \textbf{S}^{N=4}_{\textbf{2}}(r):=\begin{pmatrix}
\textbf{I} & \textbf{0} &\textbf{0} &\textbf{0}\\
  \textbf{0}& \cosh{r}\textbf{ I}&\sinh{r}\textbf{ Z}&\textbf{0}\\
    \textbf{0}&  \sinh{r}\textbf{ Z}&\cosh{r}\textbf{ I}&\textbf{0}\\
    \textbf{0} &\textbf{0} &\textbf{0}&\textbf{I}\\
    \end{pmatrix},\label{eq:bigS2}
\end{equation}
where \textbf{0} is the $2\times2$ zero matrix and, the beamsplitters simulating loss on modes $b$ and $c$ is represented by the symplectic map
 \begin{equation}
    \textbf{B$^{N=4}$}({T_b,T_c}):=\begin{pmatrix}
    \sqrt{{T_b}}\textbf{ I} & \sqrt{1-{T_b}}\textbf{ I}&\textbf{0} &\textbf{0}\\
    -\sqrt{1-{T_b}}\textbf{ I} &  \sqrt{{T_b}}\textbf{ I}&\textbf{0} &\textbf{0}\\
    \textbf{0} &\textbf{0} &  \sqrt{{T_c}}\textbf{ I} & \sqrt{1-{T_c}}\textbf{ I}\\
    \textbf{0} &\textbf{0} & -\sqrt{1-{T_c}}\textbf{ I} &  \sqrt{{T_c}}\textbf{ I}\\
    \end{pmatrix}.\label{eq:bigBS}
\end{equation}
From here, the covariance matrix for this system is calculated, assuming vacuum input modes, to be
\begin{equation}
\textbf{V}_\text{sys}=\textbf{B$^{N=4}$}({T_b,T_c})\textbf{S}^{N=4}_{\textbf{2}}(r)\textbf{I}_8(\textbf{S}^{N=4}_{\textbf{2}}(r))^T(\textbf{B$^{N=4}$}({T_b,T_c}))^T,
\end{equation}
where $\textbf{I}_8$ is the $8\times8$ identity matrix. 

For a two-mode squeezed state, the individual quadratures do not exhibit squeezing or anti-squeezing, whereas combinations of them do. Squeezing is found with $V(\Hat{q}_-)$ and $V(\Hat{p}_+)$ which are equal, while anti-squeezing is found with $V(\Hat{q}_+)$ and $V(\Hat{p}_-)$ which are also equal, where, we define $\Hat{q}_{\pm}=(\Hat{q}_b\pm\Hat{q}_c)/\sqrt{2}$ and $\Hat{p}_{\pm}=(\Hat{p}_b\pm\Hat{p}_c)/\sqrt{2}$. Expanding the equations results in 
\begin{align}
&V(\Hat{q}_{\pm})=\frac{V(\Hat{q}_b)}{2}+\frac{V(\Hat{q}_c)}{2}\pm V(\Hat{q}_b,\Hat{q}_c)\text{ and}\\
&V(\Hat{p}_{\pm})=\frac{V(\Hat{p}_b)}{2}+\frac{V(\Hat{p}_c)}{2}\pm V(\Hat{p}_b,\Hat{p}_c),
\end{align}
from which the entries of $\textbf{V}_\text{sys}$ are inserted to calculate how the two-mode squeezing is affected by channel loss. Thus, we arrive at
\begin{equation}
V(\Hat{q}_{-})=V(\Hat{p}_+)=1-\frac{T_b+T_c}{2}+\frac{T_b+T_c}{2}\cosh{2r}-\sqrt{T_b T_c}\sinh{2r}\text{ and}\label{eq:sq}
\end{equation}
\begin{equation}
V(\Hat{q}_{+})=V(\Hat{p}_-)=1-\frac{T_b+T_c}{2}+\frac{T_b+T_c}{2}\cosh{2r}+\sqrt{T_b T_c}\sinh{2r}.\label{eq:antisq}
\end{equation}

\section{Two-mode squeezing \lowercase{$r$} extraction}\label{app:powvssq}
To provide an accurate estimate of squeezing in our demonstrations, the squeezing and anti-squeezing are measured versus pump power to extract the best-fit $r$. After filtering, the free-space 771-nm pump is directly coupled into the fiber attached to the input of the PPLN waveguide. Given space-constraints and the need for a bulky thermal power meter (Coherent Model 210) for high powers, at a low power using a smaller form-factor power meter (Thorlabs S120C), the transmission from free space, coupling into fiber, through the waveguide, into the output fiber was measured to be $\eta_{W}=0.53$; these two power meters are calibrated within about 10\% of one another within their overlapping range. The best-fit $r$ is found using \eqref{eq:sq}, \eqref{eq:antisq}, $r=\sqrt{a L^2 (P_W/\eta_W) \eta_P}$~\cite{Serkland:95}, $T_b=0.3097$ (-5.09~dB), $T_c=0.2576$ (-5.89~dB), $L=2.5$~cm, $a=24$\%/W/cm$^2$, measurements of squeezing and anti-squeezing versus pump power measured after going through the waveguide $P_W$, and finally, the best fit $r$ is extracted from fitting for $\eta_P$. In this way, not only does fitting for $\eta_P$ allow finding the best fit coupling into the waveguide, but also it includes any miscalibration of the power meters or error in $a$ or $L$. Thus, an accurate fitted value of $r$ is obtained for the power used in our experiments, i.e., the highest power measured. Moreover, to allow for the squeezing and anti-squeezing to be included simultaneously in the fitting for $r$, $P\rightarrow -P$ for the squeezing data and a piece-wise function is created where for $P\leq0$ $P\rightarrow -P$ and \eqref{eq:sq}, but for $P>0$, $P\rightarrow P$ and \eqref{eq:antisq} are used. 

\begin{figure}
\centerline{\includegraphics[width=0.5\textwidth]{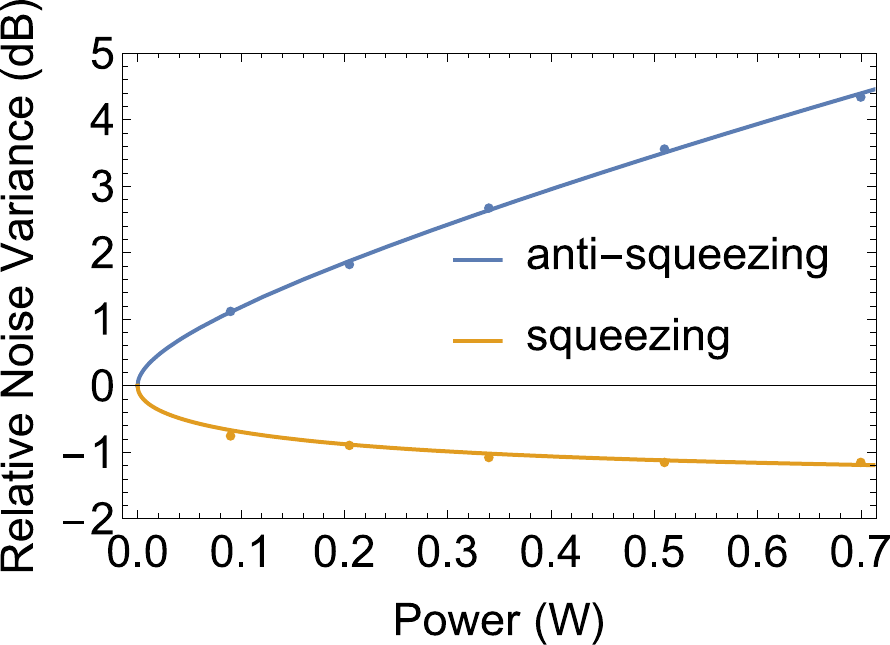}}
\caption{(anti-)squeezing versus pump power. Here the pump power is measured after the waveguide. We fit the squeezing and anti-squeezing versus power using a piece-wise function and find $\eta_P=0.49019$ with coefficient of determination R$^2$=0.9995.}
\label{fig:powvssq}
\end{figure}

Fig.~\ref{fig:powvssq} shows the measured squeezing and anti-squeezing versus the pump power measured after the waveguide. Using the piece-wise function described above, the optimal $\eta_P$ is found by fitting; the optimal fitted $\eta_P=0.49019$ with coefficient of determination R$^2$=0.9995. Using another fiber of the same type, our free-space-to-fiber coupling efficiency is measured to be about 0.82. The manufacturer measured the pump-fiber-to-waveguide coupling efficiency to be about 0.62. Thus, we estimate the coupling of the free-space pump into the waveguide to be about $\eta_P=0.51$. Our fitted value is within a few percent of the value expected but is a few percent lower than the measured transmission between free-space and the waveguide output indicating there is a bit of error in several parameters. In spite of that error, the fitting gives us an accurate value of $r$ to use for our estimation of squeezing for the different configurations that we tested.

\begin{backmatter}
\bmsection{Funding}
Advanced Scientific Computing Research (ERKJ355); U.S. Department of Energy (DE-AC05-00OR22725).
%Content in the funding section will be generated entirely from details submitted to Prism. Authors may add placeholder text in the manuscript to assess length, but any text added to this section in the manuscript will be replaced during production and will display official funder names along with any grant numbers provided. If additional details about a funder are required, they may be added to the Acknowledgments, even if this duplicates information in the funding section. See the example below in Acknowledgements.

\bmsection{Acknowledgments}
We thank Benjamin Lawrie for sharing some lab space for the deployed fiber measurements and Raphael Pooser for useful discussions. This work was performed at Oak Ridge National Laboratory, operated by UT-Battelle for the U.S. Department of Energy under contract no. DE-AC05-00OR22725. Funding was provided by the U.S. Department of Energy, Office of Science, Office of Advanced Scientific Computing Research, through the Transparent Optical Quantum Networks for Distributed Science Program (Field Work Proposal ERKJ355).
\bmsection{Disclosures}
The authors declare no conflicts of interest.

\bmsection{Data Availability Statement}
Data underlying the results presented in this paper are not publicly available at this time but may be obtained from the authors upon reasonable request.

\end{backmatter}

%%%%%%%%%%%%%%%%%%%%%%% References %%%%%%%%%%%%%%%%%%%%%%%%%
%%%%%%%%%% If using BibTeX:
%\bibliographystyle{opticajnl.bst}
%\bibliography{twomosqcoexistforjnl}

\end{document}